\documentclass[twocolumn, tighten]{aastex7_preprint}

\usepackage{lmodern, fix-cm}
\usepackage{verbatim}
\usepackage{soul}
\usepackage{newtxtext,newtxmath}

\usepackage[T1]{fontenc}
\usepackage{multirow}
\usepackage{graphicx}	
\usepackage{amsmath}	
\usepackage{amssymb}	
\usepackage[utf8]{inputenc}
\usepackage{mathtools}

\usepackage{color}      
\usepackage{placeins}
\usepackage{longtable, booktabs, threeparttablex,}

\usepackage[caption=false]{subfig}
\usepackage{blindtext}
\usepackage[normalem]{ulem}
\usepackage{subfiles}
\usepackage{hyperref}

\makeatletter
\let\oldthebibliography\thebibliography
\let\endoldthebibliography\endthebibliography
\renewenvironment{thebibliography}[1]{%
  \oldthebibliography{#1}%
  \footnotesize 
}{%
  \endoldthebibliography
}
\makeatother

\newcommand{\mi}[1]{\textsf{m12i}}
\newcommand{\mf}[1]{\textsf{m12f}}
\newcommand{\mm}[1]{\textsf{m12m}}
\newcommand{\mr}[1]{\textsf{m12r}}
\newcommand{\mc}[1]{\textsf{m12c}}
\newcommand{\mw}[1]{\textsf{m12w}}
\newcommand{\mb}[1]{\textsf{m12b}}

\newcommand{\Msol}[1]{M\textsubscript{\(\odot\)}}

\newcommand{\upenn}{Department of Physics \& Astronomy, University of Pennsylvania, 209 S 33rd St, Philadelphia, PA 19104, USA}
\newcommand{\cca}{Center for Computational Astrophysics, Flatiron Institute, 162 5th Ave, New York, NY 10010, USA}

\shorttitle{Quadrupole in the Milky Way}

\begin{document}
\thispagestyle{empty} 
\title{Shaping the Milky Way: The interplay of mergers and cosmic filaments}

\correspondingauthor{Arpit Arora}
\email[show]{arora125@sas.upenn.edu}

\author[0000-0002-8354-7356]{Arpit Arora}
\affiliation{Department of Astronomy, University of Washington, Seattle, WA 98195, USA}
\affiliation{\upenn}
\email{arora125@sas.upenn.edu}

\author[0000-0001-7107-1744]{Nicol\'as Garavito-Camargo}
\affiliation{\cca}
\affiliation{Steward Observatory, University of Arizona, 933 North Cherry Avenue, Tucson, AZ 85721, USA}
\email{jngaravitoc@arizona.edu}

\author[0000-0003-3939-3297]{Robyn E. Sanderson}
\affiliation{\upenn}
\email{robynes@sas.upenn.edu}

\author[0000-0003-2660-2889]{Martin D. Weinberg}
\affiliation{Department of Astronomy, University of Massachusetts, Amherst, MA 01003-9305}
\email{weinberg@astro.umass.edu}

\author[0000-0003-1517-3935]{Michael S. Petersen}
\affiliation{Institute for Astronomy, University of Edinburgh, Royal Observatory, Blackford Hill, Edinburgh EH9 3HJ, UK}
\email{michael.petersen@roe.ac.uk}

\author[0000-0003-3213-8736]{Silvio Varela-Lavin}
\affiliation{Departamento de F\'isica y Astronom\'ia, Universidad de La Serena, Av. Juan Cisternas 1200 Norte, La Serena, Chile}
\email{silvio.varela@userena.cl }

\author[0000-0003-4232-8584]{Facundo A. G\'omez}
\affiliation{Departamento de F\'isica y Astronom\'ia, Universidad de La Serena, Av. Juan Cisternas 1200 Norte, La Serena, Chile}
\email{fagomez@userena.cl}

\author[0000-0001-6244-6727]{Kathryn V. Johnston}
\affiliation{Department of Astronomy, Columbia University, 550 West 120th Street, New York, NY, 10027, USA}
\email{kvj@astro.columbia.edu}

\author[0000-0003-3922-7336]{Chervin F. P. Laporte}
\affiliation{LIRA, Observatoire de Paris, Universit\'e PSL, Sorbonne Universit\'e, Universit\'e Paris Cit\'e, CY Cergy Paris Universit\'e, CNRS, 92190 Meudon, France}
\affiliation{Institut de Ci\`encies del Cosmos (ICCUB), Universitat de Barcelona (IEEC-UB), Mart\'i i Franqu\`es 1, E-08028 Barcelona, Spain}
\affiliation{Kavli IPMU (WPI), UTIAS, The University of Tokyo, Kashiwa, Chiba 277-8583, Japan}
\email{chervin.laporte@icc.ub.edu}

\author[0000-0003-2497-091X]{Nora Shipp}
\affiliation{Department of Astronomy, University of Washington, Seattle, WA 98195, USA}
\email{nshipp@uw.edu}

\author[0000-0001-8917-1532]{Jason A. S. Hunt}
\affiliation{School of Mathematics \& Physics, University of Surrey, Stag Hill, Guildford, GU2 7XH, UK}
\email{j.a.hunt@surrey.ac.uk}

\author[0000-0003-0715-2173]{Gurtina Besla}
\affiliation{Steward Observatory, University of Arizona, 933 North Cherry Avenue, Tucson, AZ 85721, USA}
\email{gbesla@arizona.edu}

\author[0000-0003-3729-1684]{Elise Darragh-Ford}
\affiliation{Department of Physics, Stanford University, 382 Via Pueblo Mall, Stanford, CA 94305, USA}
\email{edarragh@stanford.edu}

\author[0000-0001-5214-8822]{Nondh Panithanpaisal}
\affiliation{Carnegie Observatories, 813 Santa Barbara St, Pasadena, CA 91101, USA}
\affiliation{TAPIR, California Institute of Technology, Pasadena, CA 91125, USA}
\email{nondh@sas.upenn.edu}

\author[0000-0003-2594-8052]{Kathryne J. Daniel}
\affiliation{Steward Observatory, University of Arizona, 933 North Cherry Avenue, Tucson, AZ 85721, USA}
\email{kjdaniel@arizona.edu}

\collaboration{all}{The EXP collaboration}
\noaffiliation 

\begin{abstract}
The large-scale morphology of Milky Way (MW)-mass dark matter (DM) halos is shaped by two key processes: filamentary accretion from the cosmic web and interactions with massive satellites. Disentangling their contributions is essential for understanding galaxy evolution and constructing accurate mass models of the MW. We analyze the time-dependent structure of MW-mass halos from zoomed cosmological-hydrodynamical simulations by decomposing their mass distribution into spherical harmonic expansions. We find that the dipole and quadrupole moments dominate the gravitational power spectrum, encoding key information about the halo's shape and its interaction with the cosmic environment. While the dipole reflects transient perturbations from infalling satellites and damps on dynamical timescales, the quadrupole---linked to the halo's triaxiality---is a persistent feature. We show that the quadrupole's orientation aligns with the largest filaments, imprinting a long-lived memory on the halo's morphology even in its inner regions ($\sim30$ kpc). At the virial radius, the quadrupole distortion can reach 1-2 times the spherical density, highlighting the importance of environment in shaping MW-mass halos. Using multichannel Singular Spectrum Analysis, we successfully disentangle the effects of satellite mergers and filamentary accretion on quadrupole. We find that, compared to isolated MW-LMC simulations that typically use a spherical halo, the LMC-mass satellite induces a quadrupolar response that is an order of magnitude larger in our cosmological halo. This highlights the need for models that incorporate the MW's asymmetry and time-evolution, with direct consequences for observable structures such as disk warps, the LMC-induced wake, and stellar tracers---particularly in the era of precision astrometry.
\end{abstract}

\section{Introduction} \label{sec:intro}

The formation and evolution of galaxies are intimately tied to the morphology of their dark matter (DM) halos, which assemble hierarchically through anisotropic accretion along cosmic filaments and mergers with smaller halos \citep[e.g.,][]{white1978core, blumenthal1984formation, springel2005simulations}. The surrounding large-scale environment dictates the angular momentum and accretion history of these halos, imprinting morphological and kinematic signatures that persist over cosmic time \citep{eisenstein1994analytical, aubert2004origin, hahn2007properties, vera2011shape}. 

The shapes of Milky Way (MW)-mass halos, in particular, are complex and evolve in response to both internal and external processes. In their inner regions $\sim$30 kpc, baryonic contraction and mergers tend to produce oblate or prolate shapes \citep{dubinski1991structure, kazantzidis2004effect, debattista2013s, bryan2013impact, vera2013constraints, bovy2016shape}. Beyond those radii, the same halos become markedly triaxial, retaining memory of their filamentary assembly and anisotropic infall \citep{allgood2006shape, vera2011shape, despali2014triaxial, valluri2021detecting, baptista2023orientations}. past major mergers, tidal interactions, and ongoing cosmic accretion continuously distort the outer halo, breaking any perfect symmetry \citep[e.g.,][]{bailin2005internal, bett2007spin, schneider2012shapes, bonamigo2015universality, gomez2017lessons, arora2020power, han2022tilt}. {A key question is whether---and how---we can recover a halo’s filamentary ``memory'' once an LMC-mass satellite has fallen in?} Most MW‑mass galaxies host or have hosted such massive companions \citep[e.g.,][]{besla2007magellanic, robotham2012galaxy}, so understanding the superposition of long‑lived filamentary distortions and transient merger signatures is crucial---both for interpreting external analogs (e.g.\ SAGA; \citealt{geha2017saga}) and for piecing together our Galaxy's own history.

The MW provides a unique opportunity to study the shape and evolution of DM halos due to the wealth of available stellar kinematic data {and to the fact that the LMC’s orbit and mass ($\sim10-20$\% of the MW; \citealt{besla2007magellanic, besla2010simulations, salem2015ram}) are exceptionally well constrained.} However, there exists a circular problem: accurate orbital modeling of stellar tracers requires a well-constrained Galactic potential, but the shape of the potential itself is inferred from the motions of these very tracers. Traditionally, the Galactic potential has been inferred by fitting stellar orbits using static, symmetric halo models \citep{binney2011galactic}. However, these assumptions are increasingly recognized as oversimplified, leading to biases in mass estimates \citep{gomez2015and, erkal2019total, vasiliev2021tango, reino2022orbital} and orbital reconstructions of stellar tracers \citep[e.g.,][]{d2022uncertainties, santistevan2024modelling}. An accurate gravitational potential is critical for (1) constraining the MW’s DM distribution---including its total mass, radial profile, and triaxiality---which remains degenerate with halo shape assumptions \citep{deason2019total, cautun2020milky}; (2) reproducing orbits of satellites and stellar streams, whose dynamics encode the nature of DM \citep[e.g.,][]{johnston2008tracing, cooper2010galactic}; (3) testing $\Lambda$CDM predictions for halo triaxiality and merger-driven evolution.

High-precision astrometric data from Gaia \citep{gaia2018gaia} have revealed that the MW’s DM halo is in disequilibrium, induced by recent massive mergers such as the LMC, with a mass of $\sim$10-20\% of the MW’s total mass \citep{besla2007magellanic}, in the outer halo \citep{garavito2019hunting, petersen2020reflex, garavito2021quantifying, petersen2021detection, conroy2021all, cavieres2025distant} and the Sagittarius dwarf galaxy in the Galactic disk \citep{johnston1995disruption, laporte2018influence, hunt2021resolving}. Recent work has incorporated time-dependent effects in MW potential models \citep[e.g.,][]{erkal2019total, shipp2021measuring, vasiliev2021tango, lilleengen2023effect}, particularly the influence of the LMC, but often assumes an initially \textit{symmetric} MW halo. However, to fully understand the global structure of the MW halo and reconstruct orbits \citep{lowing2011halo, sanders2020models, arora2024efficient, nibauer2025galactic}, it is essential to capture its intrinsic shape and evolution as well (see \citet{hunt2025milky} for a recent review).  

A key challenge in studying halo structure is disentangling the contributions of filamentary accretion and satellite perturbations, both of which distort the halo in complex, highly non-linear ways. Filamentary accretion imprints large-scale anisotropies on the halo over extended timescales, while interactions with massive satellites induce more rapid, localized distortions that propagate throughout the halo. These competing effects, along with the halo’s own response---which depends on the perturber’s mass, velocity, orbital phase, and the halo’s internal structure---create feedback that can amplify or suppress distortions \citep[e.g.,][]{gomez2016warps, laporte2018influence, garavito2019hunting}. {This raises a fundamental question for near‑field cosmology: once we subtract the well‑known LMC signals, can we still recover the halo’s primordial filamentary imprint?} Answering it demands a framework capable of systematically capturing both transient and long-lived deformations across different spatial scales. 

Basis Function Expansions (BFEs) provide a powerful approach to tackle this problem by decomposing the mass distribution into a sum of basis functions \citep{clutton1972gravitational}, often expressed in terms of spherical harmonics \citep{hernquist1992bfe, weinberg1999adaptive}. This formulation is particularly well-suited for nearly spherical systems, where perturbations naturally manifest as angular variations in the potential. In this framework, the dipole term represents the first-order deviation from symmetry, capturing large-scale displacements of the halo’s center of mass or coherent bulk flows induced by external perturbers. The quadrupole term, in contrast, reflects the halo’s intrinsic triaxial shape and/or time-dependent deformations (e.g., mergers, tidal interactions). Even in the absence of external perturbations, a triaxial halo inherently exhibits a non-zero quadrupole moment due to its anisotropic mass distribution. Higher-order terms further resolve finer structural details and localized perturbations.

BFEs have been successfully applied to model evolving halo potentials with and without LMC-mass satellites \citep[e.g.,][]{lowing2011halo, vasiliev2013new, sanders2020models, garavito2021quantifying, arora2022stability, petersen2022exp,petersen2022tidally, vasiliev23review}, allowing systematic tracking of time-dependent deformations across different spatial scales. Perturbations from massive satellites can induce long-lived dipole distortions in otherwise spherical halos \citep{weinberg2023new}. Indeed, there is a growing evidence that the MW halo exhibits a strong dipolar response to the infalling LMC, as seen in constrained MW-LMC simulations \citep{garavito2021quantifying, vasiliev2021tango, lilleengen2023effect}, cosmological zoom in simulations of MW-LMC analogs \citep[e.g.,][]{arora2024lmc, garavito2024corotation}, and even in observational data \citep{petersen2021detection, conroy2021all, yaaqib2024radial, bystrom2024exploring, chandra2024all}. Moreover, the LMC-induced dipole must be accurately modeled and subtracted before one can robustly infer the halo’s shape from stellar kinematics.

In this work, we investigate the gravitational structure of MW-mass halos from the FIRE-2 suite of zoomed cosmological hydrodynamical simulations \citep{hopkins2018fire, wetzel2023public} using time-evolving BFE models, simultaneously analyzing the temporal-spatial evolution of each harmonic. We focus on the dominant quadrupole mode of the DM halo. The quadrupole captures the overall elongation and asymmetry of the halo, encoding the cumulative impact of both its large-scale structure and dynamical response to satellites. Unlike higher-order terms, it is robust against small-scale noise while still encoding global evolution. In a companion paper \citep{darraghford2025milkyway}, we systematically compare MW-LMC analogs from the MW-est suite \citep{buch2024milky} to quantify the dipolar (bulk) response of MW-mass halos to massive satellites. Additionally, we apply a novel data-mining approach introduced in \citet{weinberg2021using} and \citet{petersen2022exp} to disentangle the effects of filamentary accretion from those induced by satellite perturbations on the quadrupole moment. This distinction is crucial for determining whether halo distortions arise primarily from smooth cosmological accretion or discrete merger-driven perturbations, thereby providing a more complete picture of halo evolution.

The paper is organized as follows: In Sec.~\ref{sec:methods}, we provide an overview of the simulations (\S\ref{sec:sims}), and basis function expansion techniques (\S\ref{sec:bfe_intro} \& \S\ref{sec:bfe_fit}), and introduce the spectrum analysis technique (\S\ref{sec:mssa_intro}). In Sec.~\ref{sec:grav_power}, we describe the gravitational power spectrum of these halos, identifying the dominant dipole and quadrupole distortions and how they evolve in time (\S\ref{sec:di_quad_time} \& \S\ref{sec:quad_evolv}). In Sec.\ref{sec:quad_fila_connect}, we quantify the connection between quadrupolar distortions and filamentary accretion, while in Sec.~\ref{sec:quad_mssa}, we separate these large-scale effects on the quadrupole gravitational structure from perturbations caused by satellite mergers. Finally, we discuss our findings and conclusions in Sec.~\ref{sec:disc_conc}.

\section{Simulations and Methods} \label{sec:methods}

In this section, we outline the key methodologies used in this work. We begin by describing the galaxy simulations utilized in the analysis (\S~\ref{sec:sims}). Next, we provide an overview of the BFE framework (\S~\ref{sec:bfe_intro}), followed by a detailed discussion of how the expansions are fitted to the simulations and an assessment of the quality of the reconstructed density profiles (\S~\ref{sec:bfe_fit}). Finally, we introduce the spectral analysis technique, used to analyze the BFE coefficient time series (\S~\ref{sec:mssa_intro}).

\subsection{Milky Way analogs} \label{sec:sims}

We use the \textit{Latte} suite of zoomed-in cosmological-baryonic simulations of isolated MW-mass galaxies, part of the Feedback In Realistic Environments (FIRE-2) project \citep{wetzel2016reconciling, wetzel2023public}.\footnote{These simulations are publicly available \citep{wetzel2023public} at \url{http://flathub.flatironinstitute.org/fire}.} These simulations employ the FIRE-2 physics model \citep{hopkins2018fire} implemented within the GIZMO code \citep{hopkins2015code}, which combines a TREE+PM solver for gravity with a Lagrangian meshless-finite-mass (MFM) hydrodynamics solver that adapts spatial resolution to local density. The FIRE-2 model incorporates detailed star formation and stellar feedback physics based on STARBURST99 stellar evolution models \citep{leitherer1999starburst99} and assumes a $\Lambda \textrm{CDM}$ cosmology consistent with Planck results \citep{collaboration2015planck}. For further details, see \citet{hopkins2018fire} for the FIRE-2 model and \citet{wetzel2023public} for the specific implementation.

The \textit{Latte} suite comprises seven simulated halos---m12b, m12c, m12f, m12i, m12m, m12r, and m12w---that resemble the MW in stellar mass, gas content, DM mass, and density profiles \citep{hopkins2018fire, garrison2018origin, sanderson2020synthetic}. These halos have total masses of approximately $1$--$2 \times 10^{12}$ \Msol{} and are resolved with initial particle masses of $m_\mathrm{b} = 7100$ \Msol{} for stars and gas, and $m_\mathrm{DM} = 35,000$ \Msol{} for DM in the zoomed-in region. Snapshots of the simulations are saved at intervals of approximately 25 Myr for the last 7.5 Gyr of evolution. While the \textit{Latte} suite focuses on isolated halos, there exists the \textit{Elvis} suite\footnote{The \textit{Elvis} suite has a higher resolution, with $m_\mathrm{b} = 3500-4000$ \Msol{} for stars and gas and $m_\mathrm{DM} = 19,000$ \Msol{} for DM, compared to the \textit{Latte} suite \citep{wetzel2023public}.} that provides paired MW-Andromeda mass simulations, which better represent the real MW environment. However, analyzing the effects of massive neighbors on the host halo's shape is beyond the scope of this work. Future studies should incorporate paired simulations to disentangle these influences.

The cosmological environments in the \textit{Latte} suite ensure realistic accretion histories, where mergers and interactions with the surrounding large-scale structure play a critical role in establishing the density distributions of the halos, assembly histories, and angular momentum profiles \citep[e.g.,][]{springel2005simulations, allgood2006shape, despali2014triaxial}. 

These zoomed-in cosmological simulations are run in a non-periodic box frame on an expanding background. Since the zoom-in regions does not have zero total momentum, we recenter the simulation onto the host galaxy frame using the shrinking spheres method \citep{power2003inner} to find the galactic center at each time step applied to star particles. Subsequently, we rotate all snapshots of the simulation to align the galactic disk with the XY plane at the present day---the principal axes. We define the host-centered rotation in the principal axes as the galactocentric frame.

We identify DM subhalos in these simulations using the {\fontfamily{qcr}\selectfont ROCKSTAR} halo finder \citep{behroozi2012rockstar}. The merger trees linking these subhalos across the simulation snapshots are constructed using the {\fontfamily{qcr}\selectfont consistent-trees} algorithm \citep{behroozi2012consistent}, as detailed in \citet{samuel2021planes}. The {\fontfamily{qcr}\selectfont consistent-trees} code reliably tracks subhalos with masses $\geq 10^6$ \Msol{} at the fiducial resolution of the FIRE-2 simulations \citep{samuel2021planes}. \citet{garavito2024corotation} present the key properties of the most massive mergers for these halos within the last 7.5 Gyr, including satellite and host masses at infall and pericenter, as well as pericentric distances. \citet{arora2024lmc} and \citet{garavito2024corotation} identify m12b as the best analog for a MW-LMC system based on its satellite-to-host mass ratio (1:8) and the orbital parameters---such as pericentric distance ($\sim40$ kpc) and orbital eccentricity (0.83)---of its most massive merging satellite.

\subsection{Characterizing halos with Basis Function Expansions} \label{sec:bfe_intro}

To analyze how the DM halo's shapes and density evolve over time in our simulations, we expand each N-body snapshot onto a spherical harmonic basis (see below). BFE provide a flexible and accurate means of capturing halo deformations \citep[e.g.,][]{vasiliev2021tango, garavito2021quantifying}, including those induced by galactic evolution and satellite mergers. They have been shown to effectively reproduce orbits, even in the presence of massive satellites, in both constrained \citep{garavito2021quantifying,lilleengen2023effect, vasiliev2024dear} and cosmological simulations \citep{lowing2011halo, sanders2020models, arora2022stability, arora2024efficient}. To model the roughly spherical DM halo shape, these expansions are constructed as bi-orthogonal functions of 3D spherical coordinates, where the angular dependence ($\theta, \phi$) is modeled using spherical harmonics, and the radial dependence ($r$) is captured by orthogonal radial functions. 

The radial functions are derived by numerically solving the Poisson equation, which is a specific form of the Sturm-Liouville equation \citep{weinberg1999adaptive}.  This approach provides flexibility in capturing arbitrary spherical profiles without the need for a parameterized zero-order model. A key advantage of this empirical basis is its ability to accurately represent both the inner and outer mass profiles of halos. Predefined basis, such as the Hernquist basis \citep{hernquist1992bfe}, suffer from force biases due to mismatched potential gradients (e.g., arising from the Hernquist basis’s $r^{-4}$ density falloff versus the $r^{-3}$ decline of NFW-like halos). Such biases are particularly critical for studies extending to the halo outskirts$\sim R_\textrm{vir}$ \citep{petersen2022exp}. By employing an empirical radial basis---free from assumptions about the density law---our method avoids these limitations, making it particularly effective in capturing the non-linear evolution of the density field, including perturbations induced by satellite mergers and filamentary accretion.

The BFE represents the density field as: 
\begin{equation}
    \rho_\mathrm{halo}(\vec{x}, t) = \sum_{\ell m n} C_{\ell m n}(t) \varrho_{\ell n}(r) Y_{\ell}^m (\theta, \phi),
\end{equation} \label{eq:bfe_dens}
where  $\varrho_{\ell n}(r)$ are the time-independent bi-orthogonal radial density functions,  $Y_{\ell}^m (\theta, \phi)$ are the spherical harmonics, and $C_{\ell m n}(t)$ are the time-dependent weighting amplitude coefficients. Here, $|m| \leq \ell$, with $\ell$ denoting the angular degree and $n$ the radial index of the polynomials. A paired potential can be derived from the density field by solving the Poisson equation\footnote{The alternative notation $\nabla^2 \Phi = 4 \pi G \rho$ is occasionally used, though not strictly accurate! Mathematically, the Laplacian operator is written as $\Delta$ which by definition is the divergence of the gradient, $\nabla^2$ can be misinterpreted as taking the gradient twice, which yields a rank-2 tensor.}, $\Delta \Phi = 4 \pi G \rho$,  where the potential $\Phi(\vec{x}, t)$ is related to the density $\rho(\vec{x}, t)$ via the Green's function. The acceleration field can then be obtained by taking the gradient of the potential, $\vec{a} = -\nabla \Phi$.

We make use of the publicly available code EXP and the associated Python library 
pyEXP \citep{exp2025JOSS} to compute and analyze the BFE of the FIRE-2 \textit{latte} suite of simulations. A comprehensive review of the BFE mathematical background was presented in \cite{petersen2022exp}.

\subsection{Fitting BFE to simulations} \label{sec:bfe_fit}

Here, we provide details on how the BFE is fit to the zoomed cosmological halos in our simulations. To analyze the shape and evolution of the DM halo and its response to infalling satellites, we focus exclusively on the smooth DM distribution of the primary halo---excluding bound substructures such as subhalos and massive satellite debris. For each snapshot, we filter out all DM particles associated with subhalos with $M_\textrm{vir} \geq 10^6$ \Msol{} from the N-body particle data. Importantly, this filtering is applied at each snapshot, meaning that as subhalos tidally strip and their contributions become part of the main halo, they are no longer excluded once the subhalo catalog loses track of them. Additionally, we remove the debris from the most massive satellite in each simulation across all snapshots. Debris removal is accomplished by identifying the peak mass snapshot of each merging satellite using merger trees (see Sec.~\ref{sec:sims}) and excluding all associated DM particles. As a result, our halos do not account for the growth driven by these most massive infalling satellites. For m12f, which has two massive satellites merging in succession, we remove the debris from both across all snapshots.  

This procedure is not without limitations. First, subhalo tracking in {\fontfamily{qcr}\selectfont ROCKSTAR} can be unreliable, especially for heavily stripped or disrupted subhalos, which may result in incomplete removal of substructure. This is particularly relevant for our most massive mergers, where tidal debris may be incorrectly classified as part of the main halo rather than the infalling satellite \citep{diemer2024haunted, mansfield2024symfind}. Additionally, by removing the anisotropically accreted debris from these satellites, we are reducing the imprint of their accretion history on the halo’s global shape. However, this choice is justified, as our goal is to isolate the intrinsic response of the MW-mass halo to the infalling satellites, rather than the persistent anisotropic distortions left by their debris.

This approach minimizes the impact of substructure, ensuring that the BFE signal is not dominated by the hard-to-model density field of small subhalos, which could introduce noise in high-order radial polynomial coefficients. It ensures that the BFE captures the halo’s intrinsic response to its environment and perturbations, while still accounting for halo growth—--except for that driven by the most massive infalling satellites. These filtered halos are directly used for the BFE fits and to compute the radial basis functions.  

We construct the radial basis functions using a spherically averaged density profile extending to 600 kpc ($\sim 2 R_\textrm{vir}$) averaged in time over the last 7.5 Gyr. This ensures that the radial functions remain fixed throughout the halo's evolution. A fixed basis is critical for studying the temporal evolution of spatial power: if the basis varied between snapshots, comparisons of structure across spatial scales would become ambiguous. In Appendix~\ref{app:radial_basis}, we demonstrate the stability of the radial basis over time, showing that it remains consistent in time except during massive mergers ($M_\textrm{sat} / M_\textrm{host} \geq 0.1 $), further justifying this approach. 


\begin{figure}
    \includegraphics[width=\linewidth]{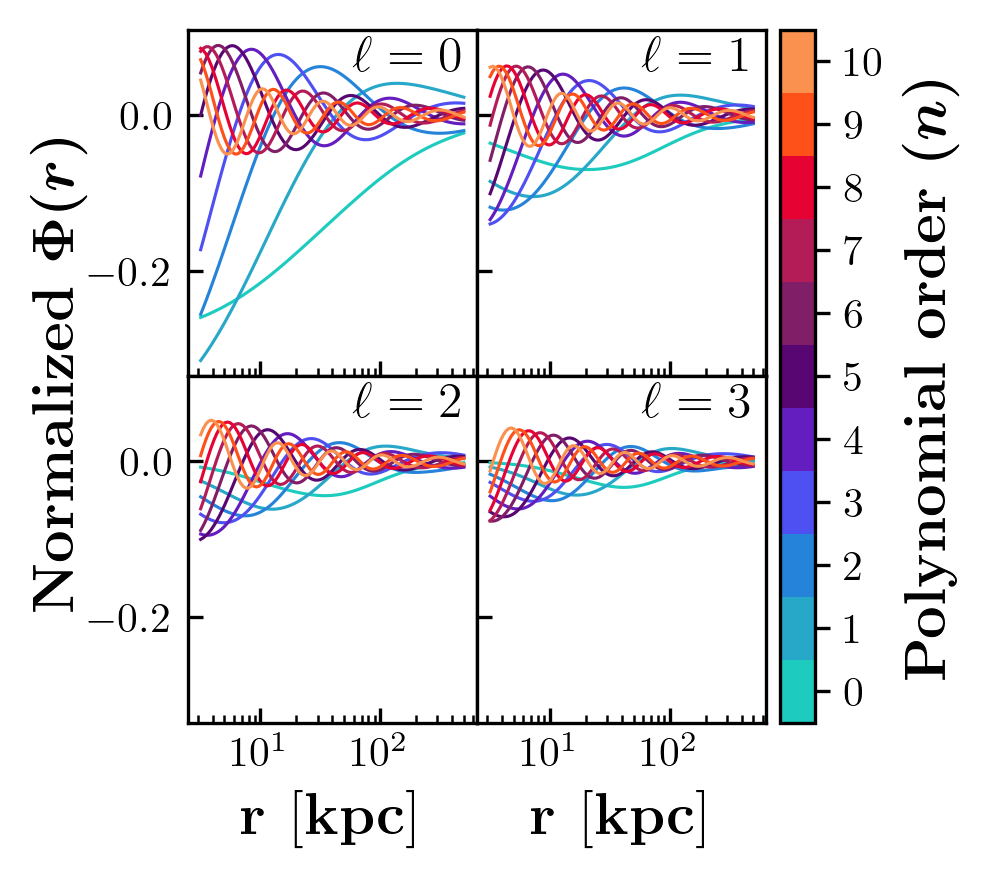}
    \caption{Orthogonal radial basis functions derived by numerically solving the Poisson equation using the time-averaged spherically averaged density profile for m12i. Each panel corresponds to a different angular degree ($\ell = 0, 1, 2, 3$), and the radial basis functions are shown as normalized potentials as a function of radial distance. Polynomials are color-coded, with redder shades indicating higher radial orders ($n$). The basis functions nodes span scales of a few kpc to hundreds of kpc, which gives resolving power for structure at all scales of interest in the problem.}
    \label{fig:basis_m12i}
\end{figure}

We truncate the expansion to $\ell_\textrm{max} \leq 4$ for the angular components and $n_\textrm{max} \leq 10$ for the radial components.\footnote{While the expansion fits are computed up to $\ell_\textrm{max} \leq20$, we later show that most of the dynamical signatures and gravitational power are contained within 4.} Fig.~\ref{fig:basis_m12i} illustrates the orthogonal radial basis constructed for m12i, with  polynomials (color-coded) shown as basis potential (normalized with ${G M_\textrm{tot}}/{r_\textrm{scale}}$) as a function of radial distance from the galactic center. Each panel corresponds to a different value of $\ell$ shown up to the octupole ($\ell = 3$), while increasing $n$ is indicated by redder shades within each panel. The basis functions satisfy bi-orthogonality conditions given by:

\begin{equation} \label{eq:biorthog}
    \int_0^\infty \varrho_{\ell n}(r) \psi_{\ell n'}(r) w(r) \, dr = \delta_{nn'},
\end{equation}

where $w(r) \equiv - 4 \pi G r^2$ is the weighting function, $\psi_{\ell n}(r)$ are the time-independent radial potential functions, and $\delta_{nn'}$ is the Kronecker delta, ensuring bi-orthogonality for fixed $\ell$. The radial basis functions span scales from a few kiloparsecs (high-order polynomials) to hundreds of kiloparsecs (low-order terms), with node spacings tuned to resolve gravitational potential variations across the full spectrum of dynamically relevant scales---from small, localized perturbations to large-scale halo asymmetries. Finally, we fit these expansions to the re-centered filtered N-body particle data for the DM halo at each snapshot for all the simulations, using particles within 600 kpc of the galactic center over the last 7 Gyr of evolution (roughly 350 snapshots).


\begin{figure}
    \includegraphics[width=\linewidth]{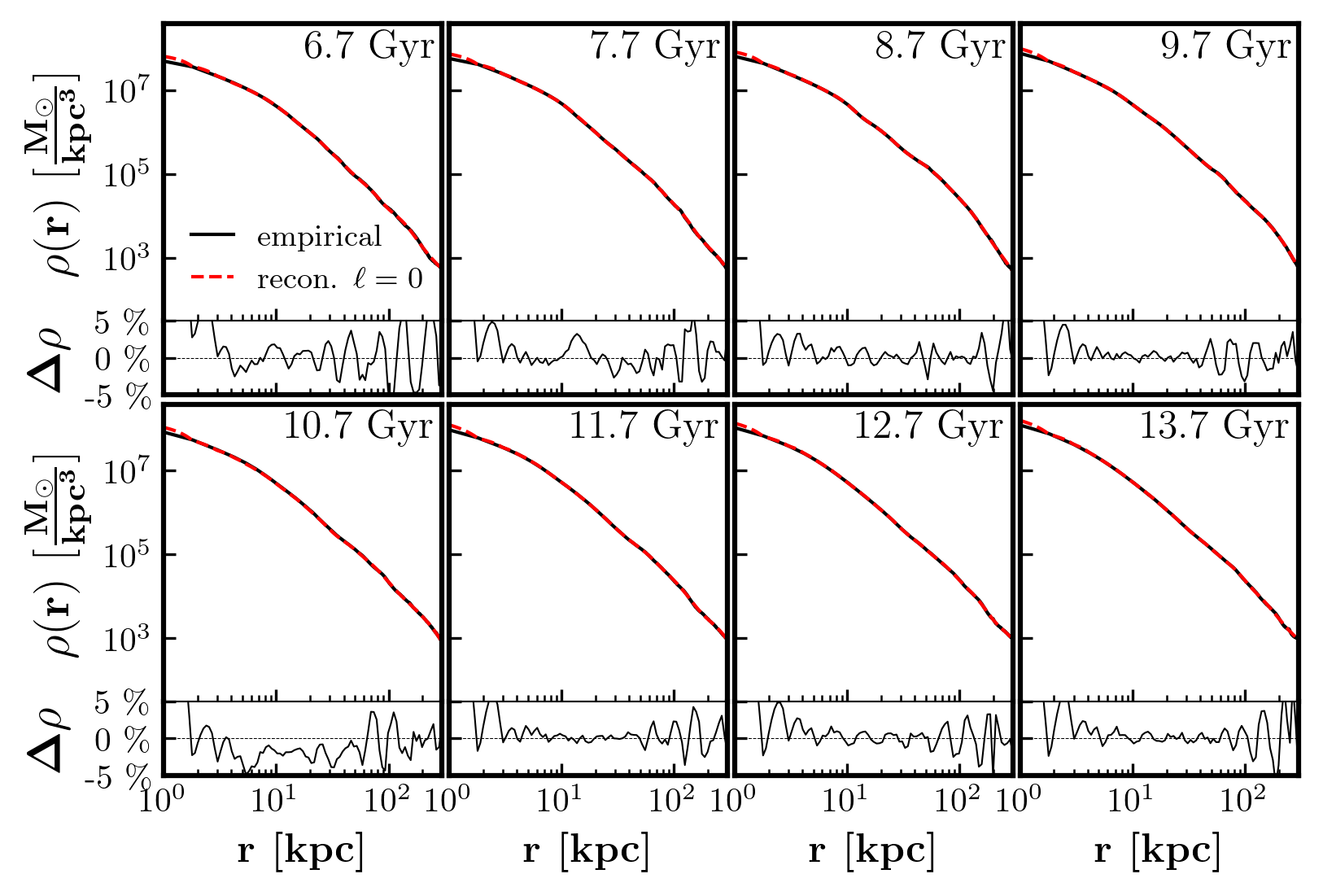}
    \caption{Spherically averaged density profiles for m12w (solid black lines) and reconstructions using the monopole ($\ell=0$) term from the BFE (dashed red lines) at 1 Gyr intervals over the last 7 Gyr. Times are labeled in the top-right corner of each panel. Subpanels show the percentage difference between the reconstructed and true profiles. Reconstructed profiles are accurate to within 5\% errors out to the virial radius at all times.}
    \label{fig:dens_sph_m12i}
\end{figure}

Fig.~\ref{fig:dens_sph_m12i} shows the spherically averaged density profile of m12w (solid black lines) and its reconstruction using the monopole ($\ell=0$) term from the BFE (dashed red lines) over the last 7 Gyr of evolution. We highlight m12w as a critical test case due to its extreme merger history: a 1:8 mass-ratio satellite with eccentricity $e=0.96$ undergoes a pericentric passage within 10 kpc at 8.7 Gyr. Such mergers induce rapid, non-adiabatic changes to the halo’s potential, challenging the fidelity of basis function reconstructions. 

The profiles are shown at 1 Gyr intervals, with the corresponding times labeled in the top-right corner of each panel. Subpanels below each main panel show the percentage difference between the reconstructed and true density profiles. Despite the merger’s disruptive effects, reconstructions are accurate to within 5\% out to the virial radius of the halo at all times. Larger discrepancies (8-10\%) are observed near the galactic center~2–3 kpc and beyond the virial radius at earlier times, reflecting the limitations of the monopole term in capturing steep central gradients and rapidly accreting outskirts. Notably, uncertainties remain stable even near the merger pericenter 8.7 Gyr, demonstrating the method’s resilience to transient perturbations.

For comparison, simulations without significant mergers (e.g., m12i), maintain uncertainties within within 5\% across all radii and times (not shown). These results align with \citet{arora2022stability, arora2024efficient}, who reported similar reconstruction errors in FIRE simulations using BFE with a grid based radial basis, even during mergers. The consistency across halo histories underscores that a fixed radial basis robustly reconstructs time-evolving density profiles, enabling reliable temporal comparisons of expansions despite non-adiabatic events. 

The time series of BFE coefficients encode rich dynamical information, including spatial and temporal correlations. Different dynamical events, such as the accretion of massive satellites or filamentary material, occur on distinct timescales and can leave imprints at specific frequencies in the BFE coefficients. Spectral analysis techniques are well-suited to disentangle these contributions and identify the dominant modes of evolution. Singular spectrum analysis (SSA) is a proven method for extracting correlated signals from time series by examining their covariance structure at successive time lags \citep{golyandina2001analysis}. By extending SSA to a multichannel framework, we can analyze the BFE coefficients---treated as multichannel time series---to uncover key modes of correlated evolution in the halo's dynamics. This approach allows us to isolate and interpret the signatures of different physical processes shaping the halo over time.

\subsection{Decomposing halo evolution and structure with mSSA} \label{sec:mssa_intro}

Multichannel singular spectrum analysis (mSSA) is a technique for identifying correlated spatial-temporal patterns across multiple time series \citep{golyandina2001analysis}. Unlike traditional SSA, which focuses on a single time series, mSSA considers multiple channels simultaneously---in our case each coefficient series $C_{\ell n m} (t)$, enabling the discovery of patterns shared across different coefficients. This makes it particularly well-suited for studying $C_{\ell n m} (t)$, which capture the dynamical evolution of a halo through their changing amplitudes. 

The method involves constructing a covariance trajectory matrix that incorporates time-lagged copies of each coefficient series over a specified window length $L$, capturing both spatial and temporal correlations. For our analysis, we adopt half the length of the time series (maximum possible length), roughly 3.5 Gyr. The advantage of the longest window is that it is sensitive to to slow trends (e.g., filamentary accretion and satellite orbits) while suppressing high-frequency noise. This effectively acts as a \textit{low-pass} filter, prioritizing signals on cosmological timescales over transient perturbations. Singular value decomposition (SVD) is then applied to this matrix, decomposing the data into orthogonal principal components (PC) ranked by their singular values. These modes represent the dominant correlated patterns, providing insight into the system's evolving structure. For a detailed description of the mathematical implementation and applications of mSSA+BFE, we refer the reader to \citet{weinberg2021using} and Appendix C of \citet{johnson2023dynamical}.

mSSA shares similarities with principal component analysis (PCA) but is distinguished by its focus on temporal patterns. While PCA identifies dominant spatial correlations from a covariance matrix built from a single field variable sampled across multiple spatial locations, SSA constructs its covariance matrix using a single time series sampled over temporal windows. The resulting temporal PCs capture the primary modes of variation across time, making SSA ideal for studying time-varying systems \citep{weinberg2023new}. Notably, mSSA generalizes PCA by incorporating temporal lags: when the window length $L=1$, mSSA reduces to PCA. For BFE coefficient series---which encode spatial structure through spherical harmonics---mSSA can be interpreted as a hybrid spatiotemporal decomposition, resolving correlated patterns across both space (via harmonic modes) and time (via lagged covariance).

mSSA has proven highly effective in recent studies. \citet{weinberg2021using} demonstrated its utility in analyzing galactic bar formation, uncovering mode coupling and pattern-speed decay, as well as a slow retrograde oscillation of the disk with a gigayear-scale period in idealized MW simulations. Similarly, \citet{weinberg2023new} used mSSA to identify dipole instabilities in NFW-like DM halos, while \citet{johnson2023dynamical} revealed two distinct disc-halo dipole modes in a MW-like galaxy: one that grows and another that decays over time, driven primarily by the halo. 

In this work, we apply mSSA to the BFE coefficients to reconstruct dynamically meaningful gravitational fields and isolate specific contributions to the halo’s morphology. In Sec.~\ref{sec:quad_mssa}, we show how mSSA disentangles the quadrupole response induced by the LMC-analog from intrinsic halo triaxiality and filamentary influences in m12b.

\section{Gravitational Power in the harmonics} \label{sec:grav_power}

The gravitational dynamics of DM halos can be studied by decomposing the density and potential fields into harmonics using BFE. This approach provides the ability to calculate gravitational power in each harmonic ($\ell$) independently, enabling a detailed exploration of the distinct dynamical features, symmetries, and phenomena captured by individual modes. Each $\ell$ harmonic encodes distinct structural and dynamical signatures of the halo. For instance, the dipole term reflects asymmetries induced by merging satellites and the motion of the center of mass \citep{weinberg1998fluctuations}, the quadrupole mainly characterizes the halo's global shape, including triaxiality and deviations from spherical symmetry \citep{vera2011shape, schneider2012shapes}. It can also result from morphological distortions associated with the resonant response of the halo to a passing satellite \citep{weinberg1998fluctuations, choi2009dynamics,d2010substructure}. Higher-order harmonics, though generally less significant in amplitude, reveal much more fine-grained features such as subhalo-induced perturbations or tidal streams \citep{lowing2011halo}.

To quantify the gravitational contributions from these different harmonic orders, we compute the gravitational power for each mode. This metric captures the spatial distribution of the gravitational field across the harmonic spectrum and provides a tool to study its evolution over time. We define the gravitational power $P=-2*W$ where $W$ is the energy in the gravitational field. For a DM halo with density, $\rho(\vec{x})$, and gravitational potential, $\Phi(\vec{x})$, $P$ is given by:

\begin{equation}
    P = - \int \rho(\vec{x}) \Phi(\vec{x}) d\vec{x}.
\end{equation}

Here, the gravitational constant is absorbed into the weighting factor of the BFE (see eq.~\ref{eq:biorthog}). In the BFE framework, the power in a specific harmonic, $P_\ell$, is calculated as:

\begin{equation}
    P_\ell = \sum_{n, m} C^2_{\ell,n, m}, 
\end{equation}

where $C_{n, m}$ are the coefficients associated with the $\ell$-th harmonic for all radial polynomial indices $n$ and azimuthal indices $m$.

Each harmonic has a distinct physical interpretation. The monopole term represents the spherically symmetric component of the density distribution, capturing the overall mass distribution and radial profile of the halo. The dipole term reflects asymmetries in the density field, often associated with the displacement of the halo center due to external perturbations or interactions with nearby structures. The quadrupole term encodes the halo's shape anisotropy and is sensitive to tidal forces, filamentary accretion, and interactions with the large-scale cosmic web. For a more detailed discussion of these modes and their implications, we refer the reader to our companion paper \citep{darraghford2025milkyway}. We note that, as we are primarily interested in the response of the main halo, we have removed debris from the most massive satellite from the particle data (see Sec.~\ref{sec:bfe_fit}). This removal systematically lowers the fractional gravitational power and asymmetry, ensuring that our analysis more realistically captures the shape and dynamics of the colder component of the main halo, leaving behind materials that is strongly unmixed.

\subsection{Halo response captured in dipole \& quadrupole}

\begin{figure}
    \includegraphics[width=\linewidth]{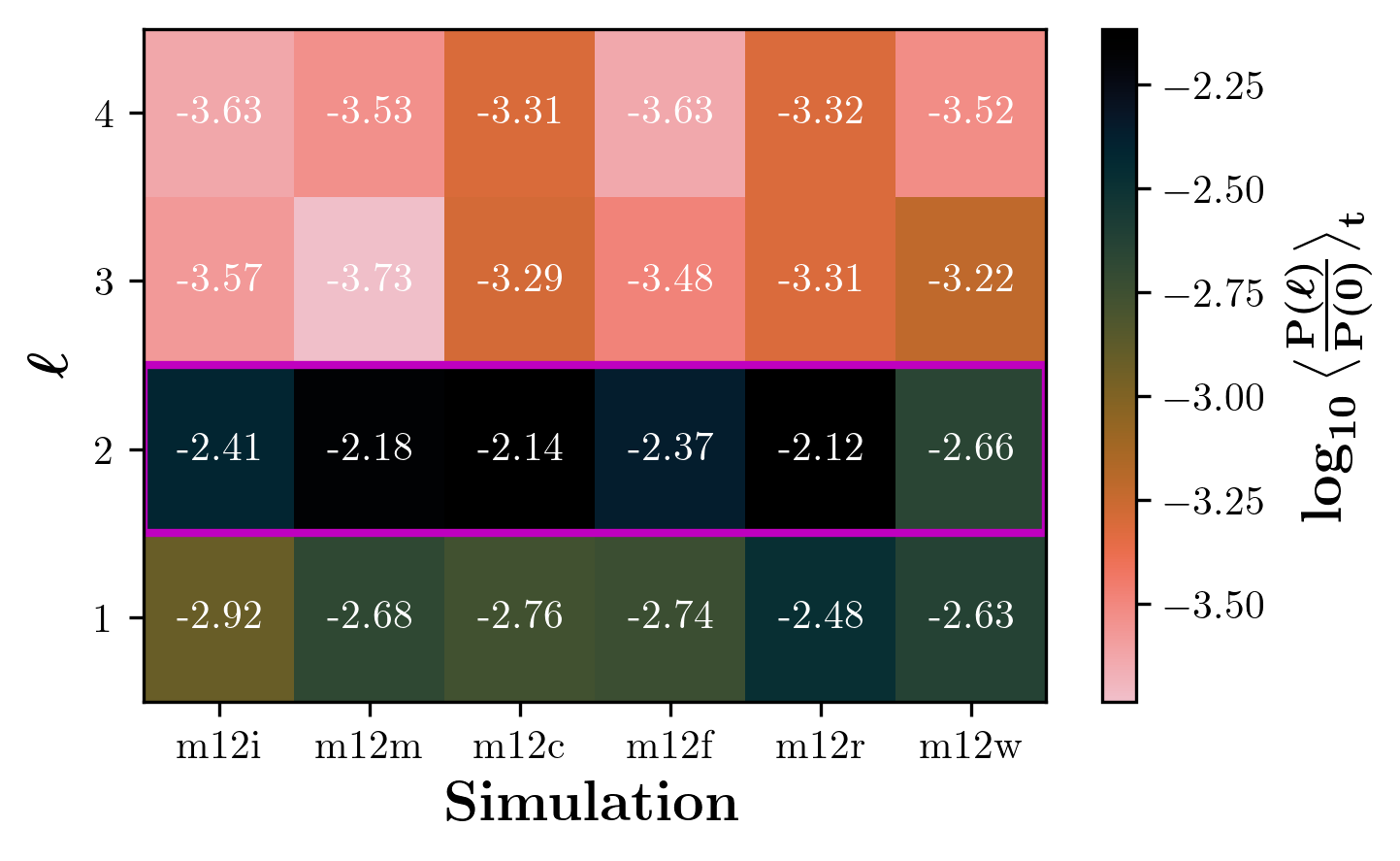}
    \caption{Power in each harmonic averaged over the last 7.5 Gyr of the halo evolution relative to the monopole ($\ell=0$) for all simulations. The largest amplitude (darker shades, and less negative values) of power are noted in the dipole ($\ell=1$) and quadrupole ($\ell=2$) modes, with the quadrupole power being systematically higher in all the simulations except \mw{}, which has a highly eccentric radial merger (with eccentricity $e=0.96$) with an LMC-mass satellite.}
    \label{fig:power_avg_poles}
\end{figure}

Fig.~\ref{fig:power_avg_poles} shows the time-averaged gravitational power in each harmonic mode ($\ell_\textrm{max} \leq 4$) relative to the monopole term, across all simulated halos (with diverse assembly histories) in the \textit{Latte} suite. The dipole ($\ell=1$) and quadrupole ($\ell=2$) consistently dominate the harmonic spectrum capturing the most relevant dynamical features of the halo, with their amplitudes almost a magnitude higher than those of the higher-order modes ($\ell 
\geq3$). These $\ell \geq3$ poles remain relatively insignificant compared to the first two harmonics. The dipole naturally arises from the motion of the halo's COM, which can be perturbed by mergers, or external tidal forces. In contrast, the quadrupole reflects the halo’s overall shape, including triaxiality and the dynamical effects of mergers and interactions that modify the halo's prolateness or oblateness. 

The $\ell=2$ systematically shows higher time-averaged power in all simulations except m12w, where the $\ell=1$ and $\ell=2$ exhibit comparable power amplitudes. The elevated $\ell=1$ in m12r and m12w is a result of an LMC-mass satellite. In m12w, this satellite reaches a pericentric distance of 10 kpc, inducing a strong asymmetric response in the halo and driving a significant increase in the $\ell=1$ power. 

Overall, the $\ell=1$ and $\ell=2$ modes capture the most gravitationally relevant contributions to the halo’s large-scale structure and dynamics \citep{quinn1992galactic, aubert2004origin, eisenstein1994analytical, vera2011shape}. These two modes are the primary drivers of the halo's response to mergers, asymmetries, and shape deviations, offering a comprehensive view of the halo's evolution. Given their dominance in the gravitational power spectrum, we focus our analysis on these two modes.

\subsection{Transient dipole \& long-term quadrupole} \label{sec:di_quad_time}

\begin{figure}
    \includegraphics[width=\linewidth]{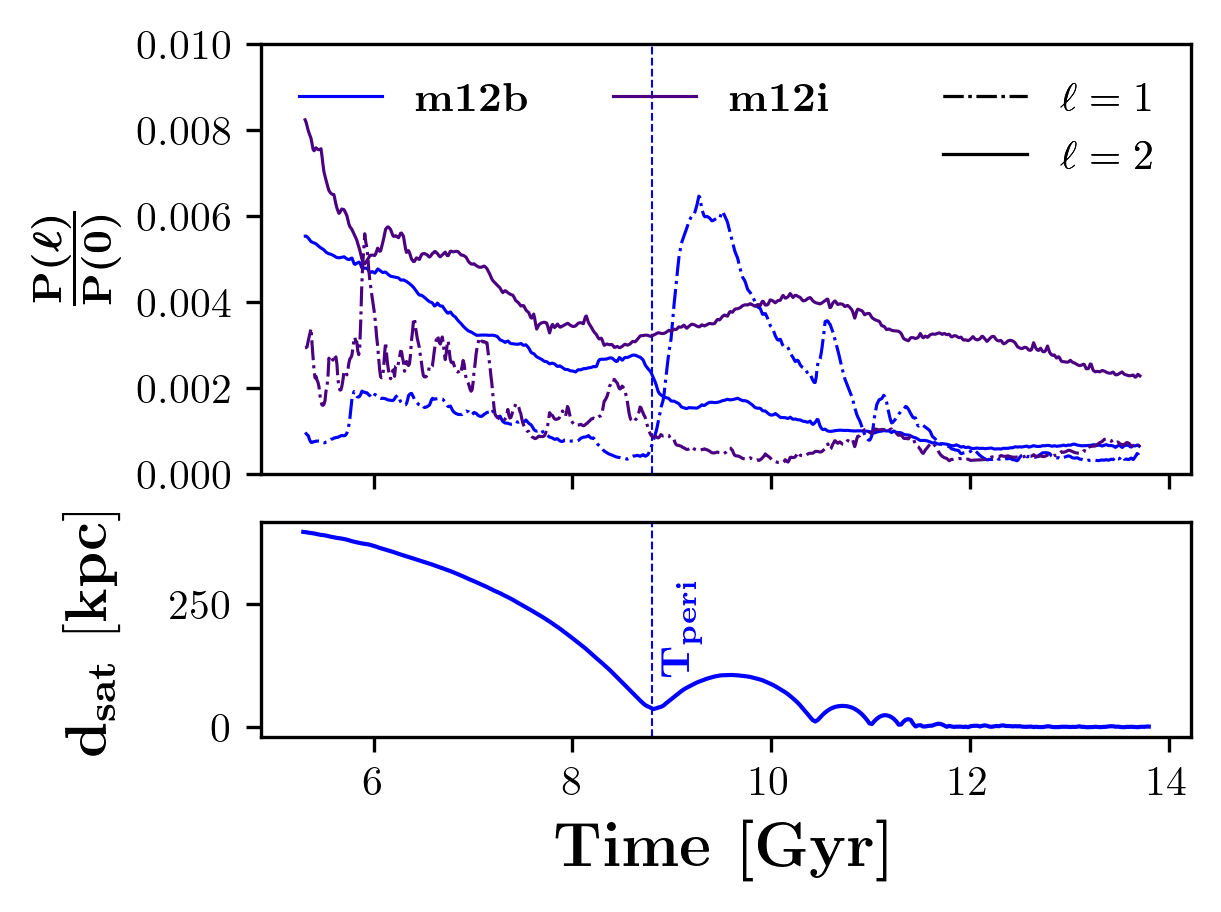}
    \caption{Top row: Fractional gravitational power as a function of time in the dipole (dotted) and quadrupole (solid) for m12b (blue, simulation with an LMC-analog) and m12i (indigo). The quadrupole power exhibits long-term evolution, with a gradual decline over time, while the dipole power shows short-term variability on merger timescales, peaking after pericentric passages. Bottom row: Total distance of the LMC-analog satellite from the halo center in m12b as a function of time. Dashed vertical lines mark the times of the first pericentric passages.}
    \label{fig:power_di_quad}
\end{figure}

Fig.~\ref{fig:power_di_quad} illustrates the temporal evolution of gravitational power in the dipole (dotted) and quadrupole (solid) modes for two simulations: m12b (blue, with an LMC-analog) and m12i (indigo). The top row shows the fractional power for each mode, while the bottom row tracks the total distance of the LMC-analog satellite from the center of the halo in m12b as a function of time. Dashed vertical line marks the time of first pericentric passage.

The dipole power exhibits transient variability on short timescales, reflecting merger-induced asymmetries. In m12b, it peaks immediately after the pericentric passage of the LMC-analog, as the halo responds to the strong perturbation. By contrast, m12i shows relatively lower dipole power after an initial peak from earlier mergers, consistent with the absence of ongoing strong interactions.

The quadrupole power, in contrast, evolves more gradually. It starts at a higher magnitude and steadily decreases over time as the halo dynamically settles. In m12b, the quadrupole also exhibits a secondary increase just before the LMC-analog’s pericentric passage, driven by the induced DM wake and torque from the satellite. This behavior highlights the quadrupole’s sensitivity to long-term shape and structural changes. We note similar trends in $\ell=1$ and $\ell=2$ power in all simulations.  

\subsection{Declining quadrupole across halo evolution}\label{sec:quad_evolv}

\begin{figure}
    \includegraphics[width=\linewidth]{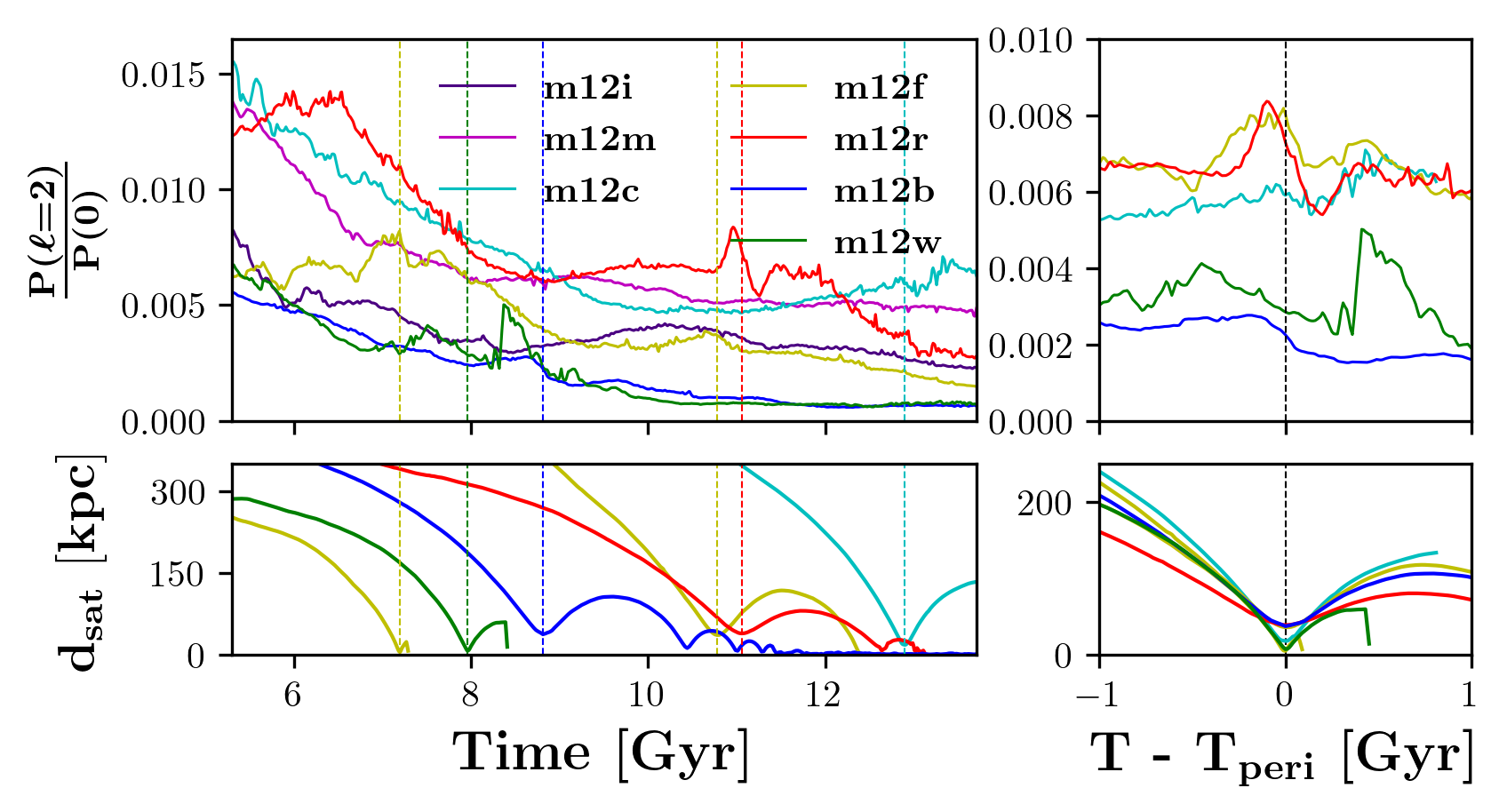}
    \caption{Top row: Fractional gravitational power in the quadrupole mode as a function of time for all simulations (color-coded) over the full 7 Gyr period (left panel) and zoomed into a $\pm1$ Gyr window around the first pericentric passage of massive satellites (right panel). Dashed vertical lines mark the times of pericentric passages. Bottom row: Distance of the most massive satellite from the halo center in each simulation (color-coded) as a function of time (left panel) and shifted to align the first pericentric passage time at 0 Gyr (right panel). The quadrupole power generally decreases as the halo evolves, but shows transient increases near pericentric passages due to tidal wakes and direct contributions from the satellites.}
    \label{fig:power_quad_all}
\end{figure}

Fig.~\ref{fig:power_quad_all} shows the temporal evolution of $\ell=2$ power across all simulations in the top row, with the corresponding orbital histories of massive satellites shown in the bottom row. The left panels provide an overview of the full 7 Gyr period, while the right panels zoom in on the $\pm1$ Gyr window around the first pericentric passage. The dashed lines indicate the times of pericentric passages for the satellites.

The quadrupole power here is primarily determined by the global shape of the halo and its triaxiality, gradually decreasing over time as the halo relaxes dynamically. This long-term decline reflects the halo's transition from an actively evolving asymmetric state to a more stable symmetric configuration (see Sec.~\ref{sec:quad_fila_connect} and Appendix~\ref{app:quad_evolv}). However, short term fluctuations in the quadrupole power are observed near the pericentric passages of satellites. These peaks are attributed to three mechanisms: (1) the dynamical friction wake induced by the satellite as it interacts with the host DM halo, and (2) the gravitational influence of the satellite, which torques the host and induces a transient quadrupole moment. (3) The satellite’s passage excites resonant oscillations in the halo’s natural modes, leading to quadrupole distortions that decay over shorter timescales. While the torque mechanism aligns with the halo’s principal axes, the resonant response arises from the halo’s natural modes being perturbed by the satellite’s gravitational field. Unlike the transient torque-driven signal, resonant oscillations can contribute to the halo’s long-term shape evolution. This distinction is critical for disentangling short-term perturbations from sustained structural changes.

Our BFE models reconstruct the halo’s density field within a spherical boundary of 600 kpc (Sec.~\ref{sec:bfe_fit}), with gravitational power computed using the full set of radial polynomials. To test sensitivity to radial truncation, we repeat the analysis for boundaries of 350 kpc and 125 kpc. While the absolute quadrupole amplitude decreases slightly at smaller scales (e.g.,~15\% at 125 kpc), the long-term decline in quadrupole power and transient merger-driven enhancements persist across all radial limits. This consistency confirms that the observed quadrupole evolution reflects intrinsic dynamical processes, not artifacts of the chosen boundary. 

To summarize, the dynamical state and shape of DM halos can be effectively studied by projecting the phase-space particle data into spherical harmonic expansions and computing their contributions to the total gravitational power. On average, the dipole and quadrupole modes dominate the harmonic spectrum. The dipole captures transient features on short timescales,  while the quadrupole represents the large-scale shape and dynamical evolution of the halo. As we demonstrate in subsequent sections, the quadrupole’s orientation and amplitude are closely tied to the halo’s interaction with filamentary structures in the cosmic environment. Although both modes can exhibit peaks due to mergers, the quadrupole power consistently is higher and decreases over time reflecting its long-term dynamical stability.

In the next sections, we will quantify how filamentary structures and satellites establish both the magnitude and orientation of the quadrupole. We will also investigate the reasons for its declining amplitude over time, providing insights into the processes that govern halo evolution and its interaction with the surrounding cosmic web.

\section{Filaments' imprint on the quadrupole} \label{sec:quad_fila_connect}

The shapes of DM halos are intrinsically set by their interactions with the surrounding large-scale structure, such as filaments and sheets of the cosmic web---a long-studied phenomenon in cosmology dating back to seminal work on structure formation \citep{hoyle1953fragmentation, peebles1969origin, doroshkevich1970space}, angular momentum acquisition \citep{white1984angular}, and anisotropic accretion \citep{eisenstein1994analytical}. Modern simulations have reaffirmed and quantified this connection, demonstrating how filaments govern angular momentum and mass accretion \citep{aubert2004origin, vera2011shape}. This alignment reflects the interplay between a halo's mass assembly history and external tidal forces, with the large-scale environment leaving a lasting imprint on the halo's morphology and quadrupole moments.

In this section, we explore the origin and evolution of the dominant quadrupole mode in these simulations. We investigate how the alignment and magnitude of the quadrupole moment correlate with the surrounding large-scale structure, specifically focusing on the filaments feeding into the halo, the weakening of these structures over time, and the evolution of the alignment.

\subsection{Filaments set the quadrupole amplitude and orientation}\label{sec:fila_orient_amp}

We briefly describe the methodologies used to measure the quadrupole position-angle and the orientations of major filaments, which enable the study of their alignment.

\begin{itemize}
    \item \textbf{Quadrole orientation ($\boldsymbol{\hat{n}_{\ell=2}}$):} The quadrupole position-angle is determined using the expansion coefficients from the spherical harmonic decomposition for $\ell=2$ and $m=0, 1, 2$ with the lowest-order radial polynomial ($n=0$). We construct the moment of inertia (I) tensor using $(C_{200}, C_{210}, C_{220})$, I's eigenvalue decomposition provides the principal axes and the position angle, denoted as $\hat{n}_{\ell=2}$. 

    \item \textbf{Filament orientation ($\boldsymbol{\hat{n}_\textrm{fila}}$):} Filament orientations are computed by identifying the location of maximum density/main filament in angular coordinates $(\theta, \phi)$ in galactocentric frame for particles associated with filamentary structures. These particles are defined as those infalling towards the galactic center (negative radial velocity), outside the virial radius of the main halo, and with particles from the main halo and subhalos removed. The orientation of the main filament is represented by $\hat{n}_\textrm{fila}$.
\end{itemize}

The position angle of the quadrupole moment ($\hat{n}_{\ell=2}$) and the filament orientation ($\hat{n}_\textrm{fila}$) allow us to quantify their alignment and explore the correlation between the two.


\begin{figure}
    \includegraphics[width=\linewidth]{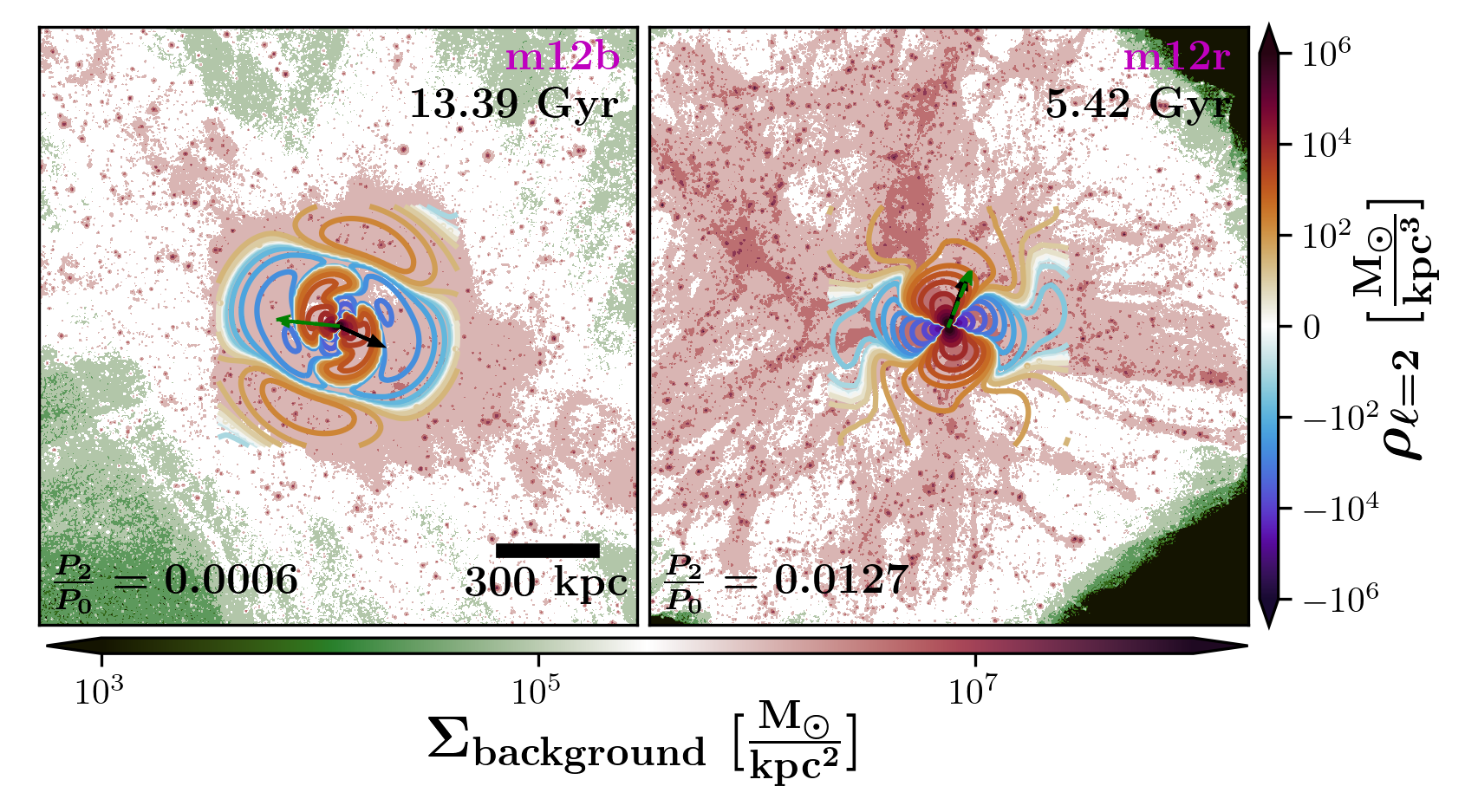}
    \caption{Quadrupole density field ($\rho_{\ell=2}$) in the XZ plane, referred to as the ``\textbf{\textit{Quadrupole-Environment View}},'' where the galactic disk is edge-on. Panels depict a time step with low fractional quadrupole power (left: m12b at 13.39 Gyr, $P_{2}/P_{0} = 0.0006$) and high quadrupole power (right: m12r at 5.42 Gyr, $P_{2}/P_{0} = 0.0127$). Red-blue open contours represent $\rho_{\ell=2}$ (positive and negative regions of the quadrupole), computed using the BFE formalism. The background shows the surface density of the environment with the main halo subtracted (diverging colormap), highlighting filamentary structures within 1 Mpc. Green arrows indicate the quadrupole orientation, $\hat{n}_{\ell=2}$, while black arrows denote the dominant filament orientation, $\hat{n}_\textrm{fila}$. In the high-power epochs, the filaments are prominent and anisotropic, and the quadrupole is strongly aligned with the dominant filament (right panel), with a 3D dot product of 0.99. In contrast, the low-power epoch corresponds to broader, less defined filamentary structures, the quadrupole power is weaker, and the alignment between $\hat{n}_{\ell=2}$ and $\hat{n}_\textrm{fila}$ is lower (3D dot product = 0.34, left panel).}
    \label{fig:fila_quad_HiLo}
\end{figure}

To explore how the large-scale environment shapes the quadrupole mode, we examine the quadrupole density field, $\rho_{\ell=2}$, in the XZ plane (the galactic disk is edge on) for two distinct epochs: one with low quadrupole power ($P_{2}/P_{0} = 0.0006$) and another with high quadrupole power ($P_{2}/P_{0} = 0.0127$). In Fig.~\ref{fig:fila_quad_HiLo}, the left panel corresponds to a low-power epoch (m12b at 13.39 Gyr), and the right panel corresponds to a high-power epoch (m12r at 5.42 Gyr).

The red-blue open contours in each panel represent the quadrupole density field, $\rho_{\ell=2}$, highlighting positive and negative contributions, computed using the BFE formalism. The background diverging colormap\footnote{Rather odd but conscious choice to highlight the structures.} shows the surface density of the environment with the main halo subtracted, revealing filamentary structures within 1 Mpc. The orientation of the quadrupole moment, $\hat{n}_{\ell=2}$, is indicated by the green arrow, while the orientation of the dominant filament, $\hat{n}_\textrm{fila}$, is shown by the black arrow. We introduce this visualization as the ``\textbf{\textit{Quadrupole-Environment View}},'' which will be used in subsequent analyses to investigate the interplay between quadrupole moments and filamentary structures across different simulations and epochs.

In the low-power epoch (left panel), the filamentary structures appear broad and diffuse as we are in a late-stage evolution of the DM halo, with weaker anisotropies in the surrounding density field. The quadrupole power is multiple orders of magnitude lower, and the quadrupole orientation is weakly aligned with the dominant filament direction\footnote{Projecting from 3D to 2D affect the apparent alignment.}, with a 3D dot product of 0.34. In contrast, during the high-power epoch at higher redshifts (right panel), the filaments are well-defined and anisotropic, and the quadrupole orientation is nearly perfectly aligned with $\hat{n}_\textrm{fila}$, with a 3D dot product of 0.99. The difference in quadrupole power between the low- and high-power epochs spans multiple orders of magnitude, underscoring the strong influence of the filamentary structures on both the amplitude and orientation of the quadrupole mode. This pattern is consistently observed across different simulations and times, reinforcing the idea that the galactic environment, particularly the filamentary structures, governs the quadrupole's behavior. 

Next, we investigate how the evolution of the filamentary environment influences and decreases the quadrupole power (Fig.~\ref{fig:power_quad_all}) and effects the overall halo shape of the halo.


\begin{figure*}[t]
    \includegraphics[width=\linewidth]{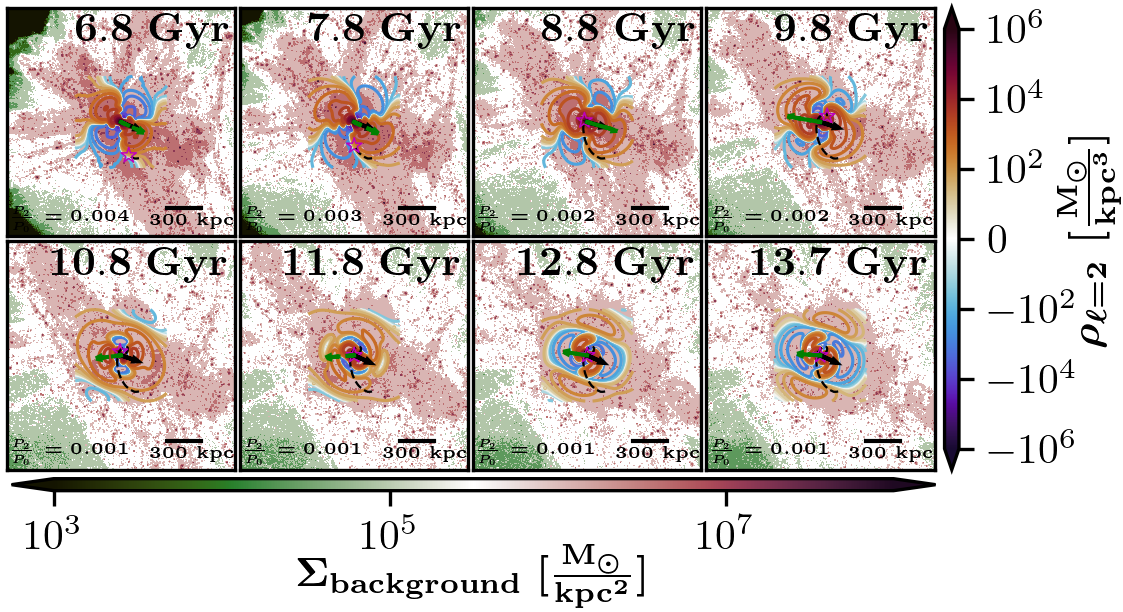}
    \caption{Time evolution of the quadrupole-environment view (as described in Fig.~\ref{fig:fila_quad_HiLo}) for the last 7 Gyr of the halo evolution. Each panel depicts a 1 Gyr time step for m12b (marked at the top of each panel). Red-blue open contours represent $\rho_{\ell=2}$ (positive and negative regions of the quadrupole) computed using the BFE formalism. The background surface density highlights filamentary structures within 1 Mpc, with the main halo subtracted. Green arrows indicate $\hat{n}_{\ell=2}$, while black arrows denote $\hat{n}_\textrm{fila}$. The trajectory of the infalling satellite is shown as a dashed black curve, with its current position marked by an unfilled magenta star. The quadrupole initially aligns with the filaments but twists and weakens as the filaments diffuse over time, and a smaller quadrupole is induced during satellite interactions.}
    \label{fig:fila_quad_evolv_m12b}
\end{figure*}

Fig.~\ref{fig:fila_quad_evolv_m12b} shows the time evolution of the quadrupole-environment view (similar to Fig.~\ref{fig:fila_quad_HiLo}) for the m12b halo over the last 7 Gyr. Each panel corresponds to a snapshot taken at 1 Gyr intervals, highlighting the evolution of the quadrupole density ($\rho_{\ell=2}$), the surrounding filamentary structures, and the trajectory of an infalling satellite. Red-blue open contours represent $\rho_{\ell=2}$, while the background colormap shows the main halo subtracted surface density of the environment. The green and black arrows indicate the orientations of the quadrupole ($\hat{n}_{\ell=2}$) and the dominant filament ($\hat{n}_\textrm{fila}$), respectively, and the dashed black curve traces the satellite's trajectory, with its current position marked by a magenta star.

At early times (e.g., the 6.8 Gyr panel), the quadrupole moment has a higher absolute density value ($\rho_{\ell=2}$) and is closely aligned with the dominant filament, as indicated by the green and black arrows in the top-left corner of the panel. Over time, as the filaments broaden and diffuse, the quadrupole twists and changes direction, while its absolute density weakens significantly. This temporal evolution demonstrates the strong connection of the filamentary structures to the quadrupole’s amplitude and orientation.

Another notable feature is the interaction between the quadrupole and the infalling satellite. Following the satellite’s first pericentric passage at around 8.8 Gyr, a secondary, smaller quadrupole appears in the inner regions, becoming especially evident near the satellite’s second pericenter at 10.8 Gyr. This interaction not only perturbs the alignment of the main quadrupole but also temporarily enhances its amplitude. 

These observations underline the importance of disentangling the quadrupole contributions from filamentary structures and satellite interactions. A detailed separation of these effects is essential for understanding how the large-scale environment and local dynamics collectively shape the quadrupole moment and, by extension, the halo’s triaxiality. In Sec.~\ref{sec:quad_mssa}, we focus on isolating these contributions and assessing their individual impacts.

Finally, we show the time evolution of the quadrupole-environment view for the remaining halos in Appendix~\ref{app:quad_evolv}. While our halos exhibit diverse configurations, such as differing numbers of filaments feeding them (e.g., m12i has two filaments, while m12r has three to four), or being embedded within massive, broad filaments (e.g., m12c and m12m), the trends observed here remain consistent. Specifically, the quadrupole weakens over time, diffuses alongside the filamentary structure, and adjusts to small perturbations caused by infalling satellites.


\begin{figure}
    \includegraphics[width=\linewidth]{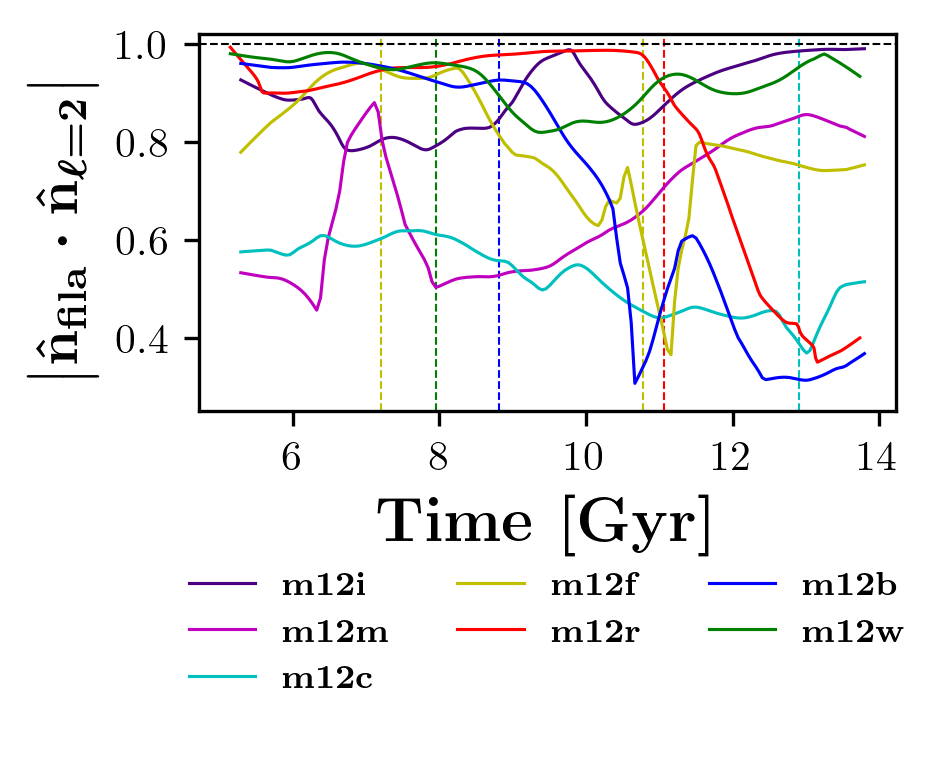}
    \caption{The absolute value of the dot product between the quadrupole position angle $\hat{n}_{\ell=2}$ and the maximum-density filament orientation $\hat{n}_\textrm{fila}$ for all simulations, shown as color-coded curves\textsuperscript{a} as a function of time. Vertical dashed lines---two for m12f---also color-coded, mark the times of first pericentric passage of massive satellites. The dot product generally remains close to 1, with variations attributable to mergers or external influences, particularly in halos embedded within massive, broad filaments such as m12m (magenta) and m12c (cyan).}\small\textsuperscript{a} {Colors are assigned based on the final letter of each halo name for consistency, e.g., m12r is red, m12i is indigo, except for m12w (green) and m12f (yellow).}
\label{fig:fila_orient}
\end{figure}

Fig.~\ref{fig:fila_orient} illustrates the temporal evolution of the alignment between the quadrupole orientation ($\hat{n}_{\ell=2}$) and the dominant filament orientation ($\hat{n}_\textrm{fila}$) for all simulations. The absolute value of their dot product is shown as color-coded curves, with each color corresponding to a specific halo (refer to Fig.~\ref{fig:power_quad_all} for the legend). Vertical dashed lines, color-matched to the respective halo, indicate the first pericentric passage of massive satellites.

The dot product remains close to 1 across most simulations, indicating a small angle between the filament and quadrupole vectors and demonstrating their strong alignment. Notable deviations occur in halos embedded within broad, massive filaments, such as m12m (magenta) and m12c (cyan). In these cases, the isotropic feeding of material from the surrounding structure weakens the alignment but does not eliminate it. For m12i (indigo), the alignment improves over time as the halo recovers from an earlier perturbation caused by a merger, ultimately achieving excellent alignment at late times.

Simulations experiencing massive mergers show characteristic drops in the dot product after the first pericentric passage, as seen in m12b (blue) and m12r (red). However, alignment often recovers in the absence of further significant perturbations, as evidenced by m12f (yellow). From Appendix~\ref{app:quad_evolv}, it is also noticed that massive satellites preferentially fall along the major filaments feeding the halo \citep[e.g.,][]{tormen1997structure, libeskind2014universal, tempel2015alignment}.

To summarize, filaments play a crucial role in setting the quadrupole moment and, by extension, the shape of the halo \citep[e.g.,][]{aubert2004origin, bailin2005internal, allgood2006shape, vera2011shape}. The strong alignment between the quadrupole and the filaments even at later times, demonstrates that the global environment has a lasting influence on the halo evolution. While satellites contribute to the quadrupole, particularly during pericentric passages, their influence is subdominant at other times compared to the persistent and dominant role of filaments in determining the overall orientation and strength of the quadrupole moment.

This raises the question of how deep in the halo the influence of filaments extend. Can filamentary structures shape the inner regions of the halo, or are their effects confined to the outskirts? Addressing these questions requires examining the radial extent of filamentary influence, as well as their role in modulating the inner halo shape, which we will do in the next section.

\subsection{Radial extent of the filaments measured with quadrupole}

\begin{figure*}
    \includegraphics[width=\linewidth]{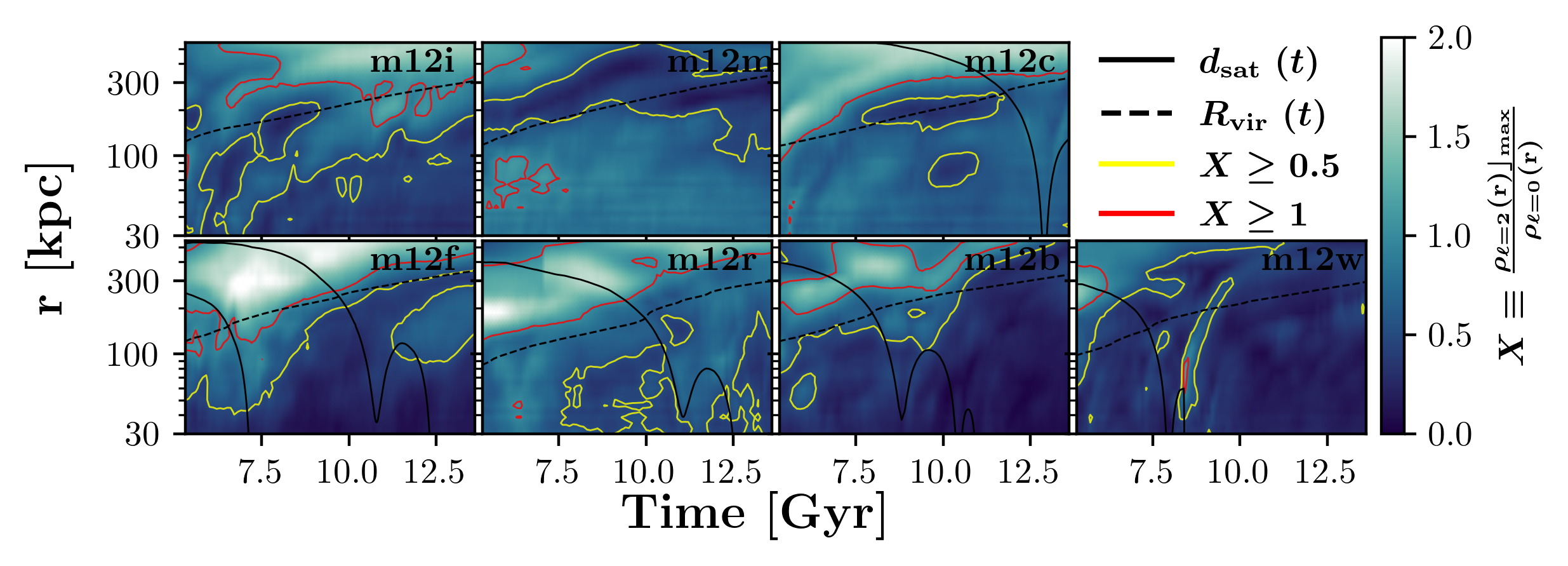}
    \caption{The maximum value of the absolute density in the quadrupole component $\rho_{\ell=2}$ at each spherical radius $r$, normalized by the spherically averaged density $\rho_{\ell=0}(r)$, shown as a function of radius and time across all simulations. The top row features simulations without massive mergers for most of their evolution, while the bottom row includes simulations with significant satellite mergers. Satellite trajectories are plotted as solid black lines, and the growing virial radius of the main halo is shown as a dashed black line. Contour levels are marked at 0.5 (yellow) and 1 (red), indicating regions of enhanced quadrupole density relative to the monopole. Fractional enhancements exceeding 1 are observed near the virial radius in most simulations, reflecting the strong influence of the large-scale filamentary environment. In all simulations, a persistent enhancement of at least 0.5 is observed as low as 30 kpc, highlighting the deep influence of the filaments on the inner halo. Localized increases in quadrupole density are also evident along satellite trajectories, corresponding to perturbations caused by mergers.} \label{fig:quad_dens_radial}
\end{figure*}

To measure the influence of the filaments throughout the halos we  compute the density of the quadrupole terms in the BFE relative to the spherical averaged density at a given radius. The fractional quadrupole density is computed as the maximum density in the quadrupole at a given radius and time, $\rho_{\ell=2}(r) \rfloor_\textrm{max}$, normalized by the spherically averaged ($\rho_{ell=0}(r, t)$) density at the same radius.  

To calculate $\rho_{\ell=2}(r) \rfloor_\textrm{max}$, we use the BFE to compute the quadrupole contribution to the density field on a uniform spherical grid with angles sampled evenly across the sphere by distributing points equidistantly along a spiral using the method described in \citet{huttig2008spiral}. This approach avoids the biased sampling inherent in traditional spherical grids (e.g., uniform $\phi$ and $\cos\theta$ grids), which disproportionately concentrate points near the poles\footnote{Traditional grids are area-uniform but angle-biased.}, leading to uneven angular resolution. The maximum quadrupole density at each radius represents the strongest aspherical contribution to the total density. The total density includes all multipoles so values of this ratio between the multiple and monopole contributions can exceed 1 for densities with strong $\ell>0$ multipoles. This fractional quadrupole density serves as a proxy for the strength of the aspherical component set by the environment of the halo at different radii and epochs.

Fig.~\ref{fig:quad_dens_radial} shows a heatmap of the fractional quadrupole density as a function of spherical radius $r$ and time for all simulated halos.  Values greater than 1 indicate regions where the quadrupole mode dominates over the spherically symmetric component, signaling highly triaxial deviations from spherical symmetry, while values near 0 correspond to spherical configurations. The top row shows simulations where the majority of the evolution occurs without massive mergers, while the bottom row depicts halos with significant satellite mergers. In each panel, satellite trajectories are marked by solid black curves, and the growing virial radius of the halo is shown as a dashed black line. Contour levels are drawn at fractional quadrupole densities of 0.5 (yellow) and 1 (red).

The $\rho_{\ell=2}(r) \rfloor_\textrm{max}$ approaches values near 1 close to the virial radius (red contours) in most halos, signaling strong anisotropy dominated by filamentary accretion. Exceptions include m12m---embedded in a single massive, broad filament that accretes matter somewhat isotropically at the virial radius---and m12w, where the primary symmetry is not captured by a quadrupole (see Appendix~\ref{app:quad_evolv}). In the other halos, early epochs ($\sim6$ Gyr, virial radius is $\sim$150 kpc) exhibit high quadrupole amplitudes ($\sim$1) at the virial radius. These amplitudes persist as the halo grows, demonstrating the retention of a ``filamentry memory'': structural imprints of anisotropic accretion that endure despite cosmic evolution. Simulations without major mergers (top row) show particularly stable alignment between the quadrupole and the virial radius, highlighting the virial radius as a dynamical transition boundary where isotropic accretion gives way to anisotropic filamentary influence. 

The inner regions ($\sim30$ kpc) in all halos retain a persistent quadrupole ( $\geq 0.5$, yellow contours), demonstrating the radial reach of filamentary influence into the inner halo morphology. However, in halos experiencing massive satellite mergers (bottom row), this filamentry memory is reconfigured by localized perturbations near the satellite trajectories. For example, in m12w, regions with higher fractional amplitudes correlate closely with the distances and times of satellite passages, particularly near pericenter, where the quadrupole enhancements are most pronounced. These perturbations form a secondary quadrupole signature superimposed on the dominant filamentary-driven mode. Importantly, the two contributions often interfere constructively, as massive satellites are typically accreted along the most massive filaments, reinforcing the overall quadrupole signal. Despite this, the satellite-induced contributions remain subdominant to the global filamentary mode. A similar increase in the amplitudes of BFE coefficients during satellite mergers was noted by \citet{arora2022stability}. 

The global influence of filaments evolves over cosmological timescales, while the localized effects of satellites are tied to their orbital dynamics. For example, Fig.~\ref{fig:power_quad_all} shows quadrupole power peaking near satellite pericenters, while Fig.~\ref{fig:fila_quad_evolv_m12b} and \ref{fig:quad_dens_radial} highlight sustained filamentary-driven quadrupole strength alongside localized increases near satellite trajectories. These distinctions in timescales allow us to disentangle the effects of mergers and filaments. In the next section, we use m12b as a case study to demonstrate how these signatures can be analyzed separately.

\section{Decoupling the effect of  filaments and satellites on the halo shape} \label{sec:quad_mssa}

In this section, we apply multichannel singular spectrum analysis mSSA to the quadrupole coefficient time series in m12b---a halo that exhibits one of the strongest transient and merger-driven quadrupole signals in our sample---to identify temporally correlated gravitational signatures. We consider all the coefficients associated with the quadrupole term, including the full set of coefficients  $(C_{\ell=2, m, n})$, where $m = 0,1,2$ and $n$ spans all the radial polynomials (Sec.~\ref{sec:bfe_intro}). The inclusion of a broad range of polynomials enables the capture of distortions across multiple radial scales, from several hundred kpc to a few kpc (see Fig.~\ref{fig:basis_m12i}). In Sec.~\ref{sec:mssa_decouple} we discuss how to decompose the competing effects from the filaments and the satellites. In Sec.~\ref{sec:mssa_density} we showed how once the PCs are identified one can reconstruct the density fields associated with the response of the halo to filament and to the satellite. 

\subsection{Spectral analysis to decouple the quadrupole response to the filaments and to the satellite}\label{sec:mssa_decouple}

mSSA operates by constructing a time-lagged covariance matrix of the input series, capturing autocorrelations within the ensemble of BFE coefficients. Each time series is detrended by subtracting its mean and normalizing by its variance to avoid amplitude bias \citep{weinberg2021using}. We use a window length $L$ corresponding to half the total time series (roughly 3.5 Gyr), to prioritize low-frequency signals tied to filamentary accretion and satellite orbital timescales. The singular value decomposition (SVD) of this matrix yields principal components (PCs), ranked by their singular values (SVs), which quantify the contribution of each PC to the total variance in the decomposition. In our analysis, the first six PCs account for $\sim 95\%$ of the total power, with the first eleven capturing $\sim 99\%$, while PCs corresponding to numerical noise remain uncorrelated and exhibit low SVs.

A key advantage of mSSA is its ability to approximate signals with time-dependent frequencies. Signals at different characteristic frequencies are projected into multiple PC groups, depending on the duration over which a given frequency persists. For instance, a persistent signal with stable frequency (e.g., filamentary accretion) will dominate fewer PCs, as its covariance structure remains coherent across the entire time series. In contrast, transient or modulated signals (e.g., satellite interactions) require more PCs to capture their time-dependent frequency shifts and localized perturbations. Consequently, a single dynamical process may span multiple PCs. By analyzing the frequency content and correlation structure of the PCs, we identify two distinct groups: one associated with filament-driven quadrupole variations (\textit{filament frequency group}) and another corresponding to the orbital response to the LMC-analog (\textit{LMC frequency group}). The detailed methodology for this classification is described in Appendix~\ref{app:mssa_pair}.

We reconstruct the BFE coefficients corresponding to the filaments and the satellite using the PC corresponding to each group.  We then compute the contribution to the quadrupole from the filaments and the LMC.  This procedure allows for a direct comparison of the disentangled filamentary and satellite-induced signatures. For details on the reconstruction procedure, see \citet{weinberg2021using} and \citet{johnson2023dynamical}.

\begin{figure}
    \includegraphics[width=\linewidth, keepaspectratio]{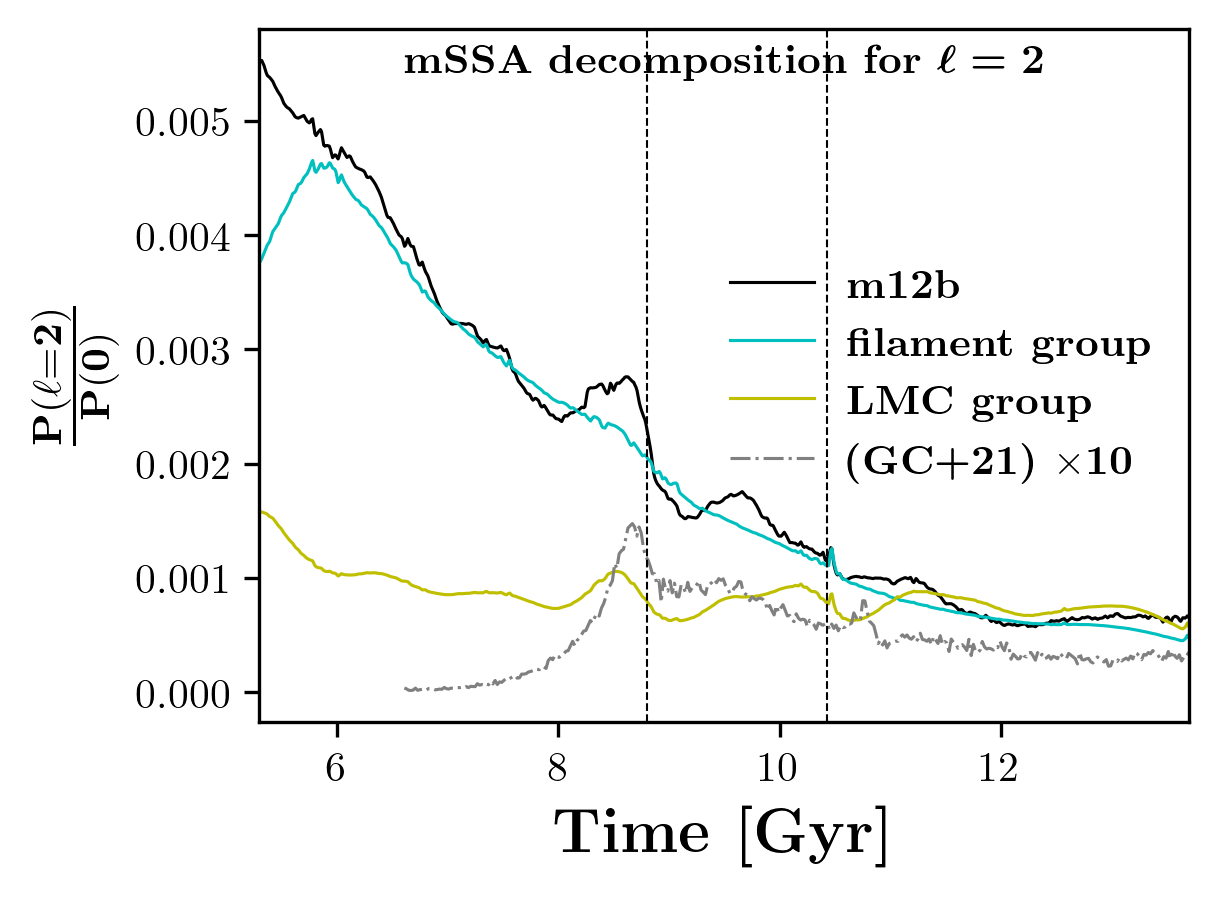}
    \caption{Reconstructed gravitational power in the quadrupole harmonics relative to the monopole as a function of time, derived from the BFE coefficients decomposed using mSSA. The cyan and yellow curves represent the contributions from the filament-associated and LMC-analog-associated modes, respectively. The dashed-dotted gray line represents the quadrupolar power in the MW-LMC constrained simulation from \citet{garavito2021quantifying}, multiplied by a factor of 10, with time shifted to align the LMC's first pericentric passage with that of the m12b LMC-analog. Vertical dashed lines indicate the first and second pericentric passages of the LMC-analog. This decomposition isolates the background halo response from the dynamical perturbations induced by the satellite.}
    \label{fig:mssa_quad_power}
\end{figure}

Fig.\ref{fig:mssa_quad_power} shows the time evolution of the quadrupole power relative to the monopole for two reconstructed components: the filament-associated (cyan) and LMC-analog-associated (yellow) quadrupole response. The total quadrupole power from the original BFE expansion is shown in black, with vertical dashed lines marking the first and second pericentric passages of the LMC-analog. The dashed-dotted gray curve shows the quadrupolar power for the MW–LMC constrained simulations from \citet{garavito2021quantifying}, multiplied by a factor of 10, with time shifted to align the LMC's first pericentric passage with that of the m12b LMC-analog.

The mSSA decomposition reveals two dominant trends in the quadrupole evolution. The first (cyan) corresponds to a long-term, gradually declining quadrupole mode driven by the filamentary structure of the halo (Sec.~\ref{sec:quad_fila_connect}). This component has a characteristic timescale of $\sim 4$ Gyr (refer to appendix~\ref{app:mssa_pair}), reflecting the slow evolution of the large-scale halo anisotropy. The second component (yellow) exhibits a shorter-timescale oscillatory pattern, with peaks in quadrupole power occurring before the pericenter passages. This mode, associated with the LMC-analog, has a timescale of $\sim 1.5$ Gyr (refer to appendix~\ref{app:mssa_pair}) and is approximately a quarter of the total quadrupole power. By the present day, both modes become comparable in magnitude, with the satellite-associated component exhibiting slightly higher power, reflecting its growing dynamical influence even after the merger.  This transient response captures the halo’s reaction to the satellite’s torque and the formation of a dynamical wake.

Notably, the amplitude of the satellite-induced quadrupole in m12b (and in other FIRE-2 analogs) is an order of magnitude higher than in MW–LMC constrained simulations (dashed-dotted gray curve), which assume an initially spherical MW halo and LMC progenitor \citep{garavito2021quantifying, lilleengen2023effect}. In those models, the development of a quadrupole mode is driven by the satellite’s orbital energy transfer, as the system first has to establish a non-spherical response. In contrast, our simulations feature a pre-existing quadrupole structure, allowing the analog’s torque to act on an already anisotropic halo, resulting in a much stronger response.

The mSSA’s ability to isolate these contributions provides a framework for precise examination of how distinct gravitational effects of large-scale structure and satellite perturbations shape the evolving halo. This will allow to compare the response across halos that are in different environments or simulated with different DM particle properties. 

\subsection{Decomposing the density field response to the filaments and to the satellite}\label{sec:mssa_density}

\begin{figure*}[ht]
    \centering
    \includegraphics[width=\linewidth, keepaspectratio]{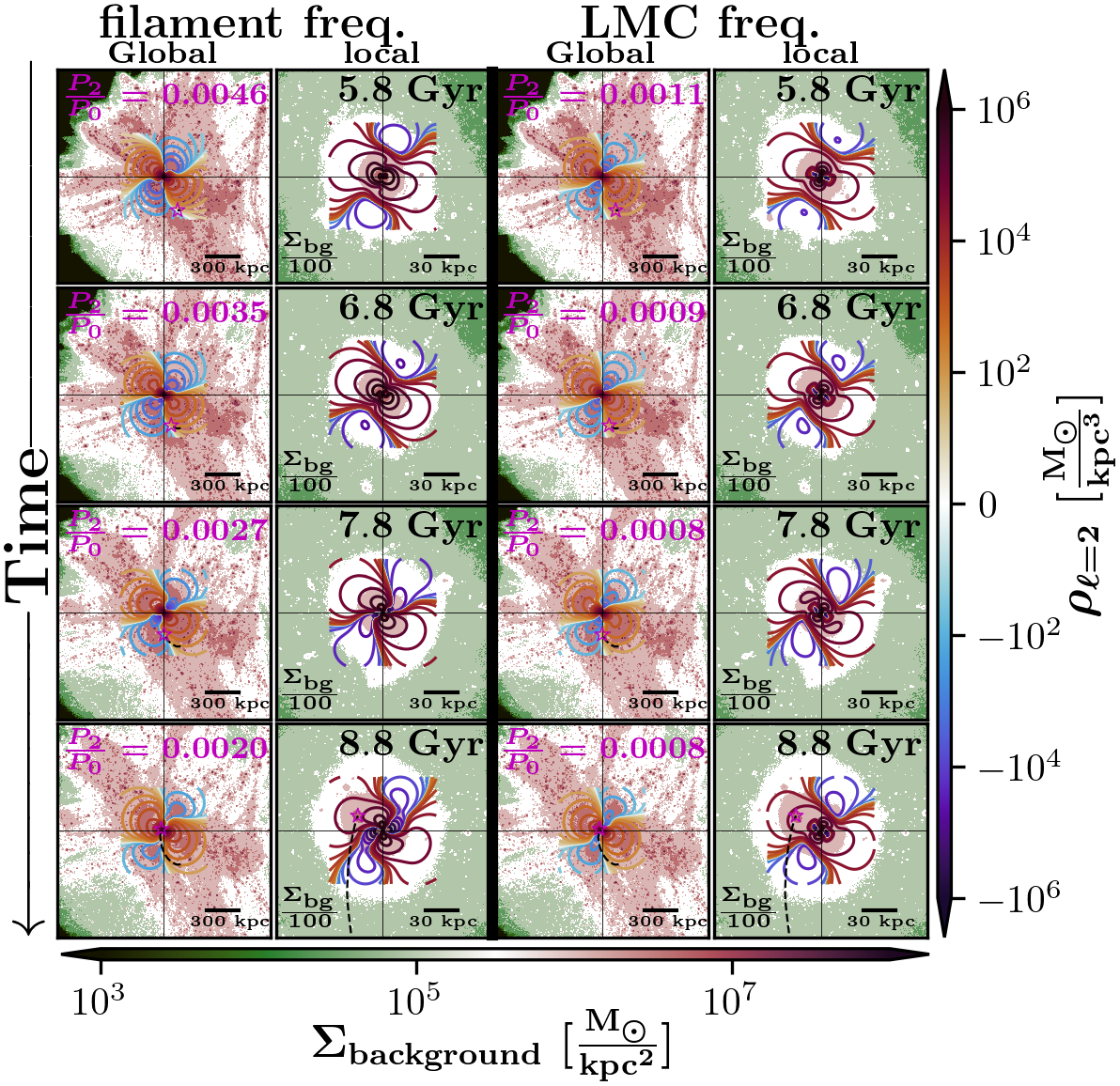}
    \caption{Time evolution of the \textit{quadrupole-environment} view (columns 1 and 3) described in Sec.~\ref{sec:fila_orient_amp}, reconstructed for the two identified frequency groups: filament frequency (first two columns) and LMC frequency (last two columns) PCs in the \mb{} halo. Each row spans from 5.8 Gyr to 8.8 Gyr. The first and third columns show the global view (within 1 Mpc), while the second and fourth columns focus on the local region (within 100 kpc). Open contours represent the quadrupole density field (vertical color bar), with the LMC-analog trajectory (dashed black curve and a magenta star for the current position) overlaid for reference. In the global view, the background highlights filamentary structures with the main halo and LMC-analog subtracted. In the local view, the background shows the main halo and LMC-analog subtracted surface densities. The reported values in the local regions are scaled down by a factor of 100 for clarity. The dominant frequency quadrupole components align with large-scale filamentary structures, while the subdominant components capture the localized impact of the satellite's passage. A full time-evolving {{\href{https://drive.google.com/file/d/12bvf-Rjiwn7rbGm4JLzInV7nr3zC9J5I/view?usp=sharing}{video}}} of this figure is available at \url{https://drive.google.com/file/d/12bvf-Rjiwn7rbGm4JLzInV7nr3zC9J5I/view?usp=sharing} and released as part of \cite{arora_dataset}}.
    \label{fig:mssa_quad_dens_0}
\end{figure*}

\begin{figure*}[ht]
    \centering
    \includegraphics[width=\linewidth, keepaspectratio]{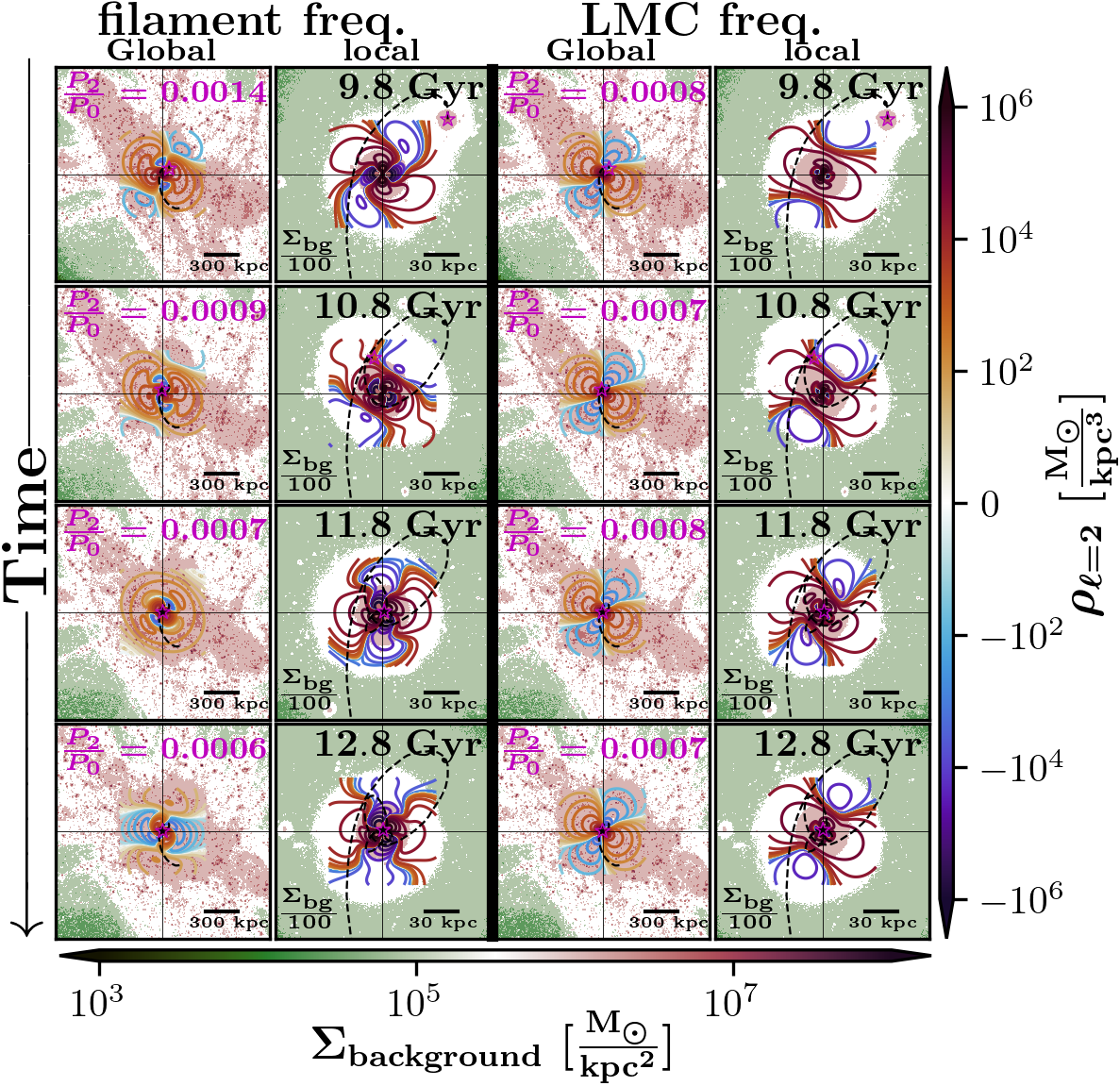}
    \caption{Same as Fig.~\ref{fig:mssa_quad_dens_0}, but for the final 4 Gyr of evolution in m12b. A full time-evolving {{\href{https://drive.google.com/file/d/12bvf-Rjiwn7rbGm4JLzInV7nr3zC9J5I/view?usp=sharing}{video}}} of this figure is available at \url{https://drive.google.com/file/d/12bvf-Rjiwn7rbGm4JLzInV7nr3zC9J5I/view?usp=sharing} and released as part of \cite{arora_dataset}}.
    \label{fig:mssa_quad_dens_1}
\end{figure*}

Once the PCs for the filaments and the satellite are identified one can compute the contribution of each PC to 
the quadrupole coefficients to obtain separate time-series expansions for the filaments and the satellite.  Fig.~\ref{fig:mssa_quad_dens_0} and Fig.~\ref{fig:mssa_quad_dens_1} illustrate the spatial distribution of the quadrupole density field at different times in the \mb{} halo, reconstructed using the two identified frequency groups from the mSSA analysis. These fields are computed by reconstructing the BFE coefficients using the PC groups associated with the filament frequency (left two columns) and the LMC frequency (right two columns). Each row represents a different snapshot in time, with Fig.~\ref{fig:mssa_quad_dens_0} covering the interval from 5.8 Gyr to 8.8 Gyr, while Fig.~\ref{fig:mssa_quad_dens_1} covers the interval from 9.8 Gyr to 12.8 Gyr of the simulation. The LMC-analog’s trajectory (dashed black curve and a magenta star for the current position) is overlaid for reference in all panels. A full video illustrating the temporal evolution of this structure is available at \url{https://drive.google.com/file/d/12bvf-Rjiwn7rbGm4JLzInV7nr3zC9J5I/view?usp=sharing} and released as part of \cite{arora_dataset}.

In each panel, the first and third columns display a global view of the quadrupole field within 1 Mpc, while the second and fourth columns provide a zoomed view of the inner 100 kpc, centered on the main halo. The quadrupole density field, reconstructed from the dominant frequency components in each PC group, is represented by open contours and is color-coded according to the vertical color bar, which denotes the 3D density values. The background in the global view panels (columns 1 and 3) highlights the filamentary structures, with the contribution from the main halo and LMC-analog subtracted to highlight the filaments. In contrast, the localized view panels (columns 2 and 4) display the surface density of the main halo with the LMC-analog subtracted, shown with a horizontal color bar. The reported density values in the local view panels are scaled down a factor of 100 for clarity. 

In the filament frequency panels, the global view (first column) reveals a strong alignment between the quadrupole structure and the surrounding filamentary environment at all times, indicating that the primary response of the quadrupole is dictated by the cosmic web. This is consistent with previous studies showing that large-scale tidal fields influence the morphology and orientation of DM halos over cosmological timescales \citep{hahn2007properties, libeskind2013alignment, codis2015spin, codis2018galaxy}. As the filaments broaden and their density diminishes over time, the power in the quadrupole component correspondingly decreases (see Sec.~\ref{sec:fila_orient_amp}), reflecting the gradual weakening of external anisotropic forces \citep{vera2011shape}. The localized filamentary view (second column) shows that the quadrupole structure in the inner halo remains largely unaffected by the presence of the LMC-analog, maintaining its alignment with the outer halo and the large-scale environment.

In contrast, the quadrupole structure reconstructed from the LMC frequency components exhibits a distinct response. The global view (third column) indicates only slight deviations in the outer halo, with the quadrupole retaining a residual memory of its initial orientation. This supports the idea that the outer shape is predominantly set by early accretion and large-scale tidal interactions, persisting long after the filamentary structures have weakened \citep{bailin2005internal, allgood2006shape}. Meanwhile, the localized LMC frequency view (fourth column) undergoes substantial torquing and modification as the analog approaches the galactic center. The density amplitude of the quadrupole field increases by an order of magnitude during the analog's passage as also seen in idealized tailored N-body simulations \citep{weinberg1998fluctuations}. Once the merger is complete and the wake is fully phase-mixed, the quadrupole stabilizes, reaching a new equilibrium state in which the local structure no longer undergoes substantial changes. However, the final alignment of the quadrupole remains distinct from that set by the filamentary structure, illustrating a non-trivial interaction between these two dominant modes that operate on different spatial and temporal scales.

\begin{figure}
    \includegraphics[width=\linewidth, keepaspectratio]{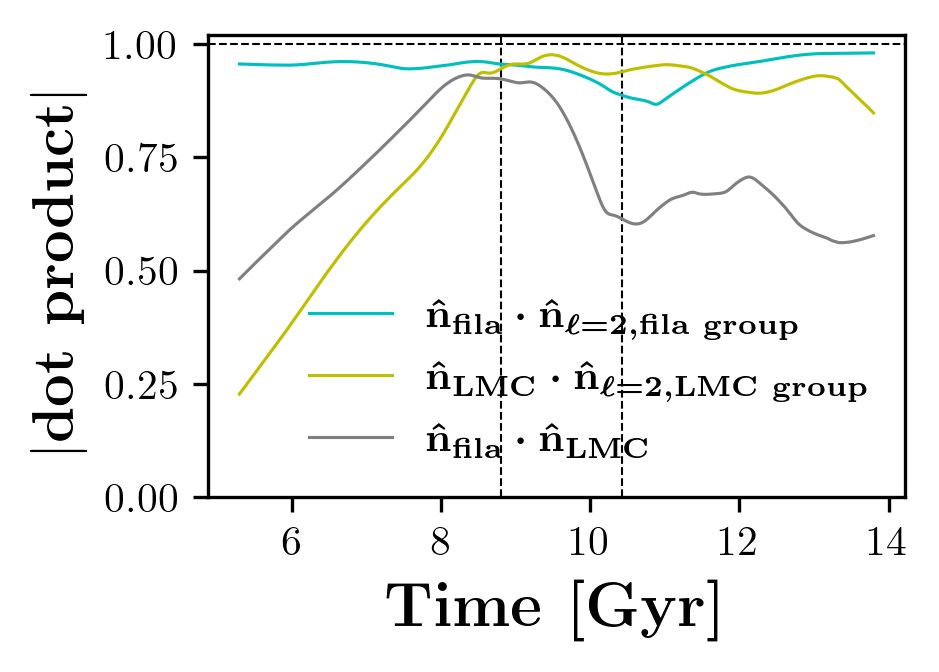}
    \caption{Temporal evolution of the absolute value of the dot product (alignment) between the PC group quadrupole position angles $\hat{n}_{\ell=2, \textrm{group kind}}$ for the filament (cyan) and LMC (yellow) groups with the maximum-density filament orientation $\hat{n}_\textrm{fila}$ and the LMC-analog position $\hat{n}_\textrm{LMC}$, respectively. The gray curve shows the alignment between $\hat{n}_\textrm{fila}$ and $\hat{n}_\textrm{LMC}$.
    Vertical dashed lines mark the first and second pericentric passages of the LMC-analog. The filament-driven quadrupole remains nearly perfectly aligned with the large-scale filaments over time, maintaining a dot product close to unity. The LMC-driven quadrupole rapidly aligns as the LMC-analog approaches first pericenter, reaching near-unity alignment and maintaining it thereafter.}
    \label{fig:mssa_quad_orient}
\end{figure}

Fig.~\ref{fig:mssa_quad_orient} quantifies the temporal evolution of the quadrupole orientation decomposed into contributions from the filamentary and LMC PC groups. The alignment here is shown as the absolute value of dot product between position angles of each component (similar to Fig.~\ref{fig:fila_orient}). The cyan curve tracks the alignment of the filament-dominated quadrupole mode with the direction of the maximum-density filament, while the yellow curve tracks the alignment of the LMC-driven quadrupole mode with the position of the LMC-analog. The gray curves shows the alignment between the maximum-density filament and LMC-analog position angles. Vertical dashed lines mark the first and second pericentric passages of the LMC-analog.

The filament-driven quadrupole remains consistently aligned with the large-scale filamentary structure, maintaining a dot product close to unity. Interestingly, we observe a slight decrease in the filament-driven quadrupole alignment immediately following the LMC-analog’s second pericenter but overall the it remains well aligned with dot product greater than 0.9. In contrast, the LMC-driven quadrupole starts not well aligned at early times, it is seen to be more aligned with the direction of the filaments (the gray curve traces the yellow curve well) and as the LMC-analog approaches first pericenter it undergoes a rapid increase in alignment, reaching near-unity and maintaining it thereafter. This suggests that the satellite’s gravitational influence reorients the local quadrupole early in its interaction, after which the alignment remains stable.

These findings demonstrate the importance of both filamentary accretion and merging satellites in shaping the structure of DM halos. The interplay between these mechanisms provides critical insights into the evolution of non-axisymmetric structures and their observable imprints in the stellar and gaseous components of galaxies. Specifically, the persistence of these structures and their alignment with the cosmic web offer a unique window into the disequilibrium dynamics of stellar halos, as observed in their kinematic and morphological signatures \citep{johnston2008tracing, pillepich2014halo, bennett2022exploring}.

In summary, we demonstrate the utility of mSSA in conjunction with BFE to decouple and analyze the time-evolving quadrupole structure of a MW-mass halo. Our decomposition identifies two dominant components: a long-term, slowly evolving quadrupole mode aligned with the filamentary large-scale structure, and a transient, satellite-induced mode associated with the LMC-analog. The filament-driven quadrupole gradually weakens over time as the filaments fade away, while the satellite-induced quadrupole undergoes significant short-term amplification and torquing during the LMC-analog’s orbital evolution. The ability to isolate these components provides a clearer framework for understanding how global tidal fields and local satellite interactions shape halo morphology, with implications for the persistence of non-axisymmetric features in galaxies.

\section{Summary and discussion} \label{sec:disc_conc}

Understanding the DM halo shape and evolution is critical for both galaxy formation theory and galactic dynamics. From a galaxy formation perspective, the structure of halos is shaped by filamentary accretion and merging satellites, which regulate how gas collapses to form galaxies, influencing star formation, angular momentum buildup, and feedback processes. Our results provide a direct measurement of how these large-scale structures imprint long-lived distortions in the halo, which in turn affect the evolution of galaxies embedded within them.

On the other hand, galactic dynamics relies on accurate halo models to reconstruct the gravitational potential. The assumption of smooth, static, and spherical halos---implicitly adopting a shared principal-axis frame with the stellar disk---is commonly used to model the orbits of stars, satellite galaxies, and stellar streams. However, our findings challenge this assumption by demonstrating that asymmetric perturbations are an inherent and persistent feature of halos, which may systematically bias dynamical inferences of DM distributions. Indeed, both observations \citep{han2022tilt, nibauer2025galactic} and simulations show that DM halos are frequently tilted by $>20^{\circ}$ relative to their stellar disks---an shown in TNG \citep{han2023tiltedTNG} and FIRE-2 Latte suite (used in this work) \citep{baptista2023orientations}---further underscores the need to go beyond aligned, spherical models when interpreting dynamical data.

In this work, we have decomposed zoomed cosmological MW–mass halos into time‑evolving spherical harmonic expansions to quantify their large‑scale gravitational distortions. Two modes capture nearly all large‑scale halo distortions (see Fig.~\ref{fig:power_avg_poles}). The dipole encodes large-scale displacements of the halo's center of mass, while the quadrupole measures the halo’s global triaxial shape and its slow evolution under filamentary accretion. Our key conclusions are:

\begin{itemize}
    
    \item \textbf{\textit{Filaments-driven quadrupole anchors halo morphology.}} The quadrupole is set early by Mpc‑scale filaments and aligns with them due to anisotropic accretion except during major mergers (see Fig.~\ref{fig:fila_orient}). Even as filaments dissipate, the halo retains a ``memory'' of its orientation (see Fig.~\ref{fig:fila_quad_evolv_m12b} and Fig.~\ref{fig:fila_quad_all})---up to 50\% density distortion at 30 kpc and 1–2× near the virial boundary (see Fig.~\ref{fig:quad_dens_radial}). The quadrupole amplitude declines gradually over several Gyr (see Fig.~\ref{fig:power_quad_all}).
    
    \item \textbf{\textit{Dipole and quadrupole amplitudes reach maxima near pericenter.}}  
    The host's dipole peaks sharply at each satellite pericenter---reaching fractional density perturbations up to unity (Fig.~\ref{fig:power_di_quad})---and damps on orbital timescales (e.g., 1-2 Gyr) and this bulk motion must be removed to recover the true intrinsic halo triaxiality (quadrupole). The quadrupole exhibits localized boosts from dynamical friction wakes and direct torquing prior to the satellite's pericenter (Fig.~\ref{fig:power_quad_all}), superimposed on a secularly declining baseline quadrupole tied to the halo's shape and cosmic environment.   

    \item \textbf{\textit{Halo response at pericenter.}}, and this bulk motion must be removed to recover the true intrinsic quadrupole.  The quadrupole then shows localized boosts from wakes and torques (Fig.~\ref{fig:power_quad_all}) on top of a long-lived baseline.


    \item \textbf{\textit{The Milky Way's present-day halo shape and response.}} Since the LMC crossed its first pericenter, both the dipole and quadrupole amplitudes in our Galaxy’s halo likely remain near their peak values---consistent with idealized $N$‑body simulations of the MW–LMC system. Accurately modeling and subtracting this ongoing dipolar motion is therefore essential before one can robustly infer the halo’s intrinsic shape.

    \item \textbf{\textit{The satellite‑induced response is boosted in cosmological halos.}} In our zoomed cosmological halos, an LMC‑mass satellite induces a quadrupole power at pericenter that is an order of magnitude larger than in constrained simulations (see Fig.~\ref{fig:mssa_quad_power}) that assume a spherically symmetric MW \citep{garavito2021quantifying, lilleengen2023effect}. This satellite-induced quadrupole response is fundamentally sensitive to the true shape of the halo (see Fig.~\ref{fig:mssa_quad_dens_0} and Fig.~\ref{fig:mssa_quad_dens_1}).

    \item \textbf{\textit{mSSA disentangles filamentary and satellite quadrupole.}} mSSA on the quadrupole coefficients cleanly isolates the slow, filament-driven quadrupole from the fast, satellite-driven response (see Fig.~\ref{fig:mssa_quad_power} and Fig.~\ref{fig:mssa_quad_orient}). Models that assume spherical halos fail to capture this superposition, systematically misrepresenting the dynamical evolution of triaxial MW-mass halos.

\end{itemize}

Accounting for the MW’s full assembly history and triaxiality---captured by our time-evolving quadrupole---has direct consequences for modeling our Galaxy’s observable structures. Triaxial halo asymmetries and past mergers can seed vertical disturbances in the disk, such as warps, corrugations, and bending waves \citep{weinberg1998fluctuations, weinberg2006magellanic, gomez2013vertical, gomez2016warps, laporte2018influence, laporte2019stellar, hunt2021resolving, han2023tilted, grand2023ever}. Notably, the larger satellite response in an asymmetric halo can exert a stronger gravitational torque on the disk, potentially resolving the discrepancy in which constrained MW–LMC simulations yield disk warps that are too small---despite accurately reproducing their spatial locations \citep{laporte2018influence, stelea2024milky}.

In the stellar halo, observed LMC-induced wake amplitudes \citep{conroy2021all, amarante2024mapping, cavieres2025distant} exceed predictions from constrained DM-only spherical halo simulations \citep{garavito2019hunting}. The triaxial halos observed in cosmological simulations enhance these wake amplitudes due to stronger gravitational coupling between the satellite and halo, naturally explaining the observed excess. Moreover, halo tracers such as satellites and stellar stream orbits are widely used to constrain the Galactic potential and the LMC’s total mass \citep{johnston1999tidal, patel2018estimating, malhan2019constraining,  erkal2019total, shipp2021measuring, koposov2023s}. While recent stream-based models incorporate the LMC’s time-dependent influence within a symmetric MW halo \citep{erkal2019total, shipp2021measuring, lilleengen2023effect}, our work shows that neglecting the halo triaxiality introduces systematic offsets in inferred masses. Including this asymmetry \citep{vasiliev2021tango, nibauer2025galactic} is essential for accurate orbit modeling and Galactic potential recovery.

The upcoming decade of deep photometric and kinematic data from Rubin/LSST \citep{ivezic2019lsst} and the Roman Space Telescope \citep{spergel2015wide} will push well beyond \emph{Gaia}’s reach. These facilities will map halo tracers with precise proper motions and radial velocities out to---and beyond---the MW’s virial radius. With this unprecedented data set, we will have the opportunity to directly test whether these asymmetries manifest in the orbits of distant halo tracers. By comparing kinematics to predictions from static versus dynamically evolving halo models, we can take the next step toward fully characterizing the morphology of the MW’s DM halo and its response to cosmic accretion.

\begin{acknowledgments}
A.A. and R.E.S. were supported in part by Simons Foundation grant 1018462, NSF grants AST-2007232 and AST-2307787, and NASA grant 19-ATP19-0068. N.G.-C. and K.J.D. acknowledge support provided by the Heising-Simons Foundation grant \# 2022-3927. R.E.S. and M.D.W. were supported in part by grant no. NSF PHY-2309135 to the Kavli Institute for Theoretical Physics (KITP). R.E.S. was also supported in part by a Sloan Fellowship. M.S.P. is supported by a UKRI Stephen Hawking Fellowship. F.A.G. acknowledges support from the FONDECYT Regular grant 1211370, by the ANID BASAL project FB210003, and from the HORIZON-MSCA-2021-SE-01 Research and Innovation Programme under the Marie Sklodowska-Curie grant agreement number 101086388. K.V.J. was supported by Simons Foundation grant 1018465. C.F.P.L. acknowledges funding from the European Research Council (ERC) under the European Union’s Horizon 2020 research and innovation program (grant agreement No. 852839). J.A.S.H. acknowledges the support of a UKRI Ernest Rutherford Fellowship ST/Z510245/1. G.B. is supported by NSF CAREER award AST-1941096 and NASA ATP award 80NSSC24K1225. The EXP collaboration gratefully acknowledges the support of the Simons Foundation.

A.A. and N.S. acknowledge support from the DiRAC Institute in the Department of Astronomy and the eScience Institute, both at the University of Washington. N.G.-C. and K.J.D. respectfully acknowledge that the University of Arizona is on the land and territories of Indigenous peoples. Today, Arizona is home to 22 federally recognized tribes, with Tucson being home to the O’odham and the Yaqui. The University strives to build sustainable relationships with sovereign Native Nations and Indigenous communities through education offerings, partnerships, and community service.

A.A. would like to thank the FIRE collaboration, the anonymous referee, and the Scientific Computing Team at the Flatiron Institute for access to data, computing resources, and valuable discussions that shaped this work. A.A. also expresses sincere gratitude to ChatGPT for its language assistance (and for writing its own acknowledgment).
\end{acknowledgments}

\software{IPython \citep{ipython}, Numpy \citep{numpy}, Scipy \citep{scipy}, {Gizmo Analysis} \citep{2020ascl.soft02015W}, pyEXP \citep{petersen2022exp, exp2025JOSS}, Rockstar \citep{behroozi2012rockstar}, {Halo Analysis} \citep{2020ascl.soft02014W}, Matplotlib \citep{matplotlib}, CMasher \citep{cmasher}.}

\appendix
\section{Temporal stability of radial basis functions} \label{app:radial_basis}

\begin{figure*}
    \centering
    \includegraphics[width=0.95\linewidth]{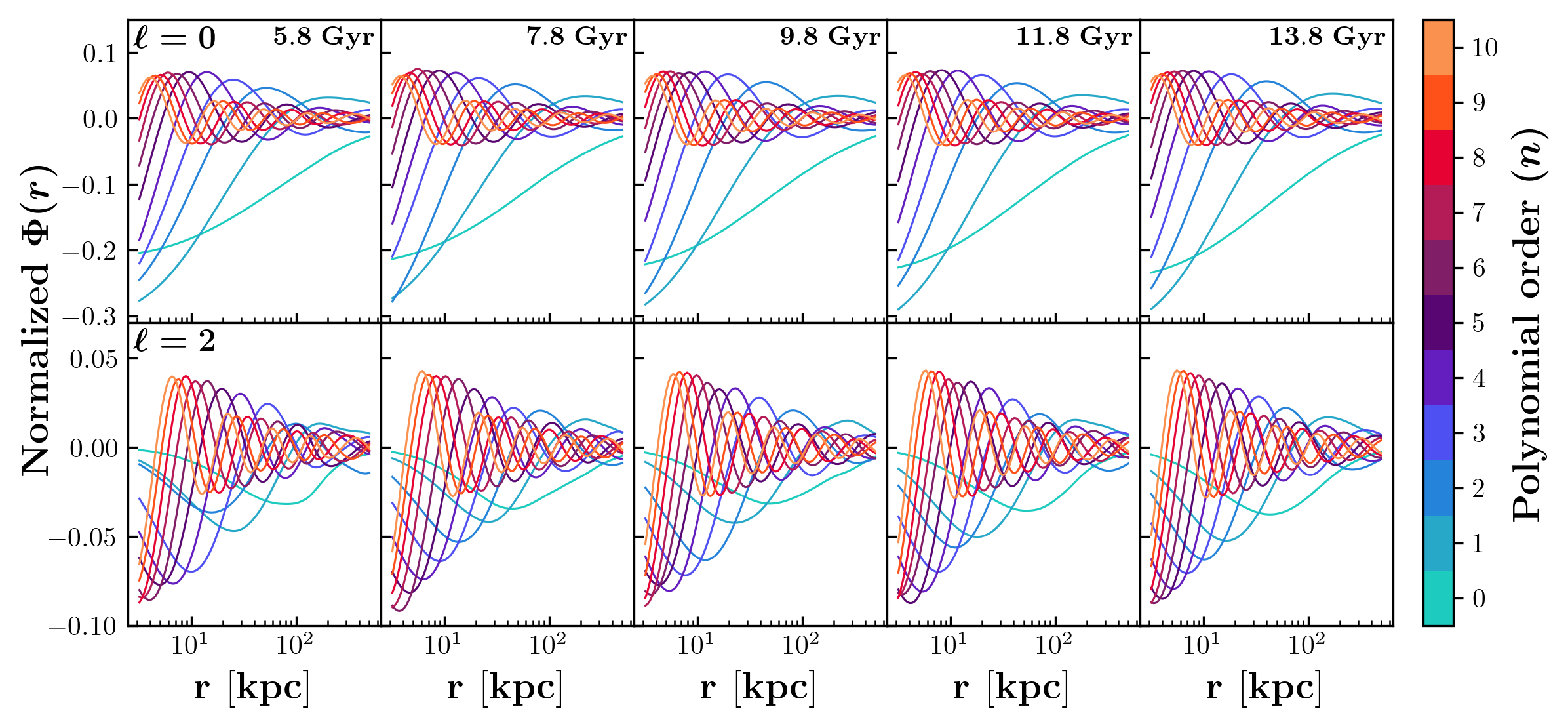}
    \caption{Temporal evolution of orthogonal radial basis functions for m12i, computed by solving the Poisson equation using the empirical density profile at each time step. Columns represent different epochs from 5.8 Gyr to 13.8 Gyr, while rows correspond to the monopole (top) and quadrupole (bottom) radial functions. The basis functions are shown as normalized potentials as a function of radial distance, color-coded by radial order ($n$), with redder shades indicating higher-order polynomials. The overall shapes and amplitudes of the basis functions remain stable over time, with minor variations in the inner regions due to the deepening of the potential. These results justify the use of a time-averaged spherical density profile to construct a fixed basis for the expansion, ensuring consistency in time-series analysis of the expansion coefficients.}
\label{fig:basis_evolv}
\end{figure*}

In this appendix, we demonstrate that the choice of a time-averaged density profile for constructing the radial basis functions is well justified. Specifically, we show that the radial polynomials computed at each time step remain largely unchanged over the course of the simulation, reinforcing the validity of using a fixed, time-averaged basis.

A full mathematical description of the construction of adaptive basis functions and the solution to the Sturm–Liouville equation (SLE) can be found in \citet{weinberg1999adaptive}. In brief, these basis functions are derived by numerically solving the Poisson equation for an orthogonal potential-density pair, using the empirically derived spherically averaged density profile of the halo at each time step. This procedure ensures that the basis functions capture the structure of the gravitational potential in a dynamically consistent manner.

To assess the stability of the basis, we compute the radial basis functions at individual snapshots and compare them over time. Fig.~\ref{fig:basis_evolv} illustrates the resulting radial basis functions for the monopole (top row) and quadrupole (bottom row), computed from 5.8 to 13.8 Gyr at 2 Gyr intervals (columns). The functions are plotted as normalized potentials, color-coded by radial order ($n$), with redder shades representing higher-order polynomials.

We observe that the overall shapes and amplitudes of the polynomials remain remarkably stable over time. While minor variations are present in the inner regions owing to the deepening of the potential with ongoing accretion, these changes occur gradually on 2–4 Gyr timescales and do not significantly alter the global structure of the basis. This stability supports our approach of using a mean density profile to construct a fixed set of radial basis functions, as the basis functions computed at any given snapshot are highly similar to those derived from the time-averaged profile (see Fig.~\ref{fig:basis_m12i}). 

An alternative approach to mitigate fluctuations in the empirical density profile is to fit the profile with a smooth double power-law function and derive the basis functions from this fitted model. This method further stabilizes the basis, particularly in cases where transient features introduce short-lived variations in the empirical density profile. However, we find that even without such smoothing, the empirical density profile itself produces basis functions that are highly stable over time.

Across all our simulated halos, we find that the basis remains stable, even in the presence of minor fluctuations. However, during pericentric passages of massive satellites—such as in m12w—the basis exhibits transient deviations. Even in these cases, the overall structure of the basis is retained outside of these brief perturbations.

Thus, using a time-averaged density profile to derive a fixed basis is a well-motivated and robust choice. It ensures that the expansion coefficients are evaluated consistently over time, and can be analyzed through mSSA to reconstruct physically meaningful gravitational structures.

\section{Quadrupole-Environment evolution for all simulations} \label{app:quad_evolv}


\begin{figure*}
    \centering
    \includegraphics[width=0.75\textwidth]{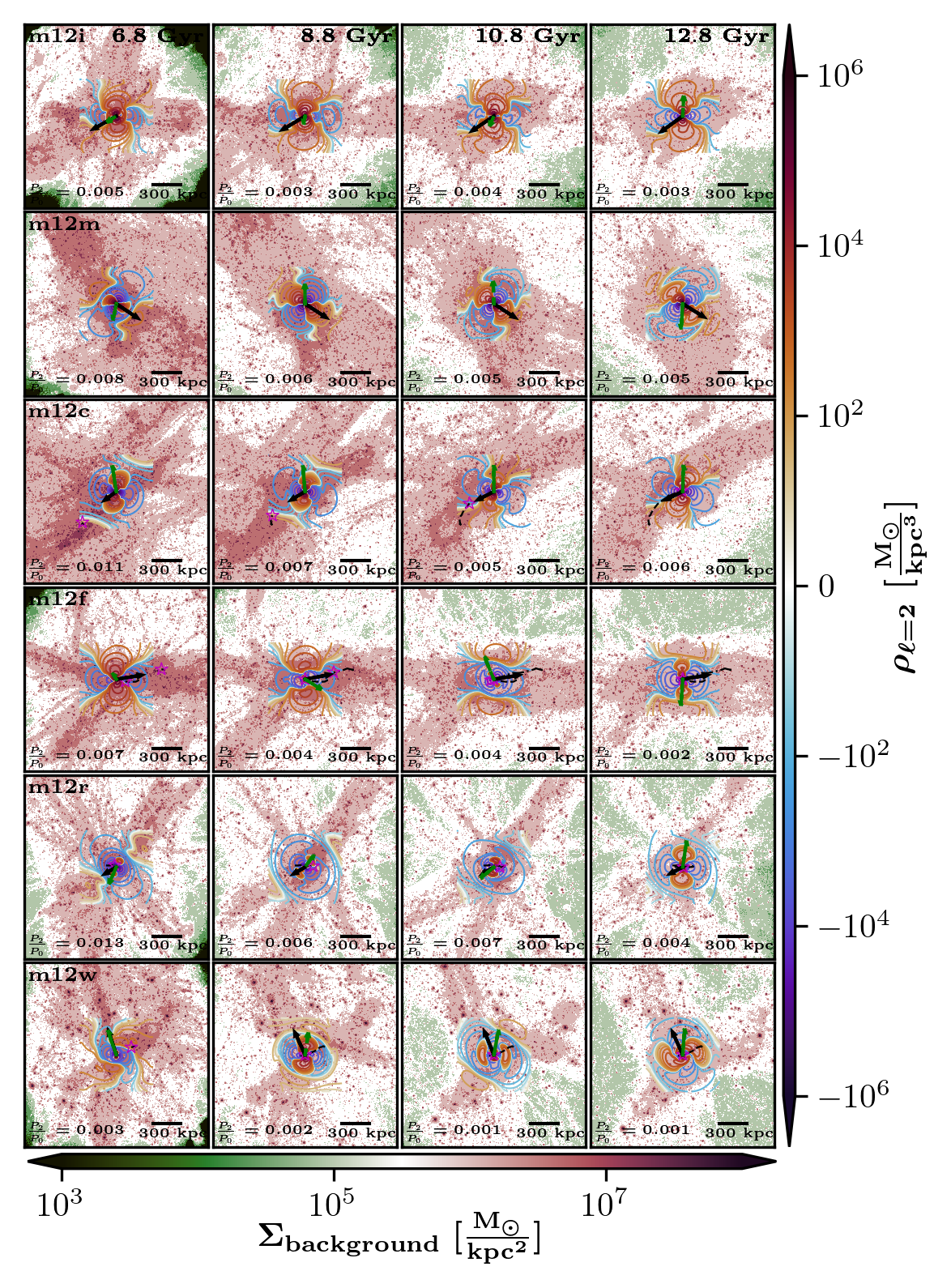}
    \caption{Time evolution of the quadrupole-environment view (as described in Fig.~\ref{fig:fila_quad_HiLo}) for the past 7 Gyr of the halo evolution. Each row corresponds to a different simulation, and each column shows a snapshot at 2 Gyr intervals, from 6.8 Gyr to 12.8 Gyr. The quadrupole density strength, $\rho_{\ell=2}$, is represented by red-blue open contours, indicating positive and negative regions, respectively, computed using the BFE formalism. The background surface density (diverging color map) highlights filamentary structures within 1 Mpc, with the main halo subtracted. Green arrows indicate the orientation of the quadrupole moment, $\hat{n}_{\ell=2}$, while black arrows denote the dominant filament orientation, $\hat{n}_\textrm{fila}$. Dashed black lines trace the trajectories of massive infalling satellites, with their current positions marked by open magenta stars. The fractional quadrupole gravitational power is marked in lower left corner in each panel. The time evolution illustrates the strong correlation between the halo quadrupole and its filamentary environment, as well as the influence of massive satellites.}
\label{fig:fila_quad_all}
\end{figure*}

Fig.~\ref{fig:fila_quad_all} presents the time evolution of the quadrupole-environment view for all simulations, complementing the examples shown in Sec.~\ref{sec:fila_orient_amp} and Fig.~\ref{fig:fila_quad_HiLo} (view description) and the detailed evolution highlighted for m12b in Fig.~\ref{fig:fila_quad_evolv_m12b}. Each row represents a different simulation, while columns show snapshots at 2 Gyr intervals over the past 7 Gyr of evolution. The corresponding fractional gravitational power in the quadrupole is marked in the lower left corner.  

We note that the simulations exhibit a wide variety of filamentary environments. For e.g., m12i is influenced by two dominant filaments, while m12r features 3–4 distinct anisotropic filaments feeding the halo. In contrast, m12m is embedded in a massive, broad filament that affects the halo nearly isotropically, whereas m12c is surrounded by a single dominant filament. Despite these differences, the amplitude of the quadrupole moment decreases consistently as the filamentary structures broaden and their surface densities decrease over time.

The evolution of the quadrupole moment varies depending on the filamentary configuration. In m12m, the gradual fading of the surrounding broad filament corresponds to a significant reduction in the quadrupole density amplitude. In m12r, the persistent presence of multiple distinct filaments maintains a stronger and more dynamic quadrupole structure over time. These trends emphasize that the halo’s quadrupole responds directly to changes in its filamentary environment, with alignment and amplitude varying based on the filaments' density distributions and spatial configurations.

Interactions with infalling satellites also contribute to variations in the quadrupole, particularly near their pericentric passages. For e.g., m12b and m12r show temporary deviations in quadrupole alignment and amplitude during such events. However, these effects are secondary compared to the dominant influence of the filamentary environment; the halo tends to retain the memory of its original quadrupole orientation \citep[e.g.,][]{dubinski1991structure, bailin2005internal, allgood2006shape, vera2011shape} even after satellite perturbations. Notably, massive satellites predominantly infall along major filaments \citep[e.g.,][]{tormen1997structure, aubert2004origin,libeskind2014universal, tempel2015alignment} , further reinforcing the connection between these structures and the halo's evolution. 

These results confirm and extend the findings in Sec.~\ref{sec:fila_orient_amp}, demonstrating that the primary driver of halo shape and evolution is the global filamentary structure feeding the halo. While satellite interactions introduce localized perturbations, the filamentary environment predominantly determines the long-term evolution of the quadrupole and, by extension, the triaxiality of the halo.

\section{Finding mSSA principal component groupings} \label{app:mssa_pair}


\begin{figure*}
    \centering
    \includegraphics[width=0.88\linewidth]{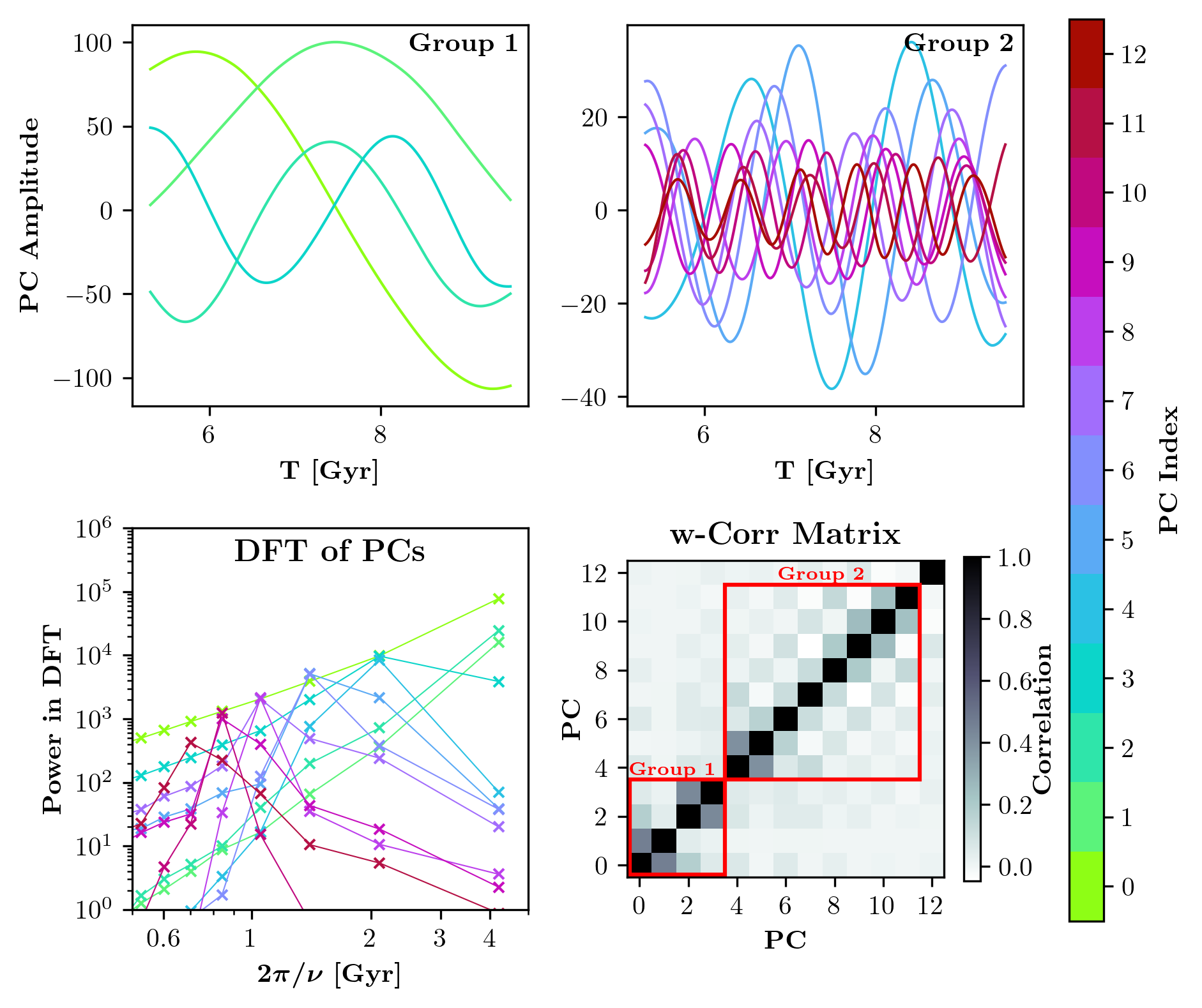}
    \caption{Principal Component (PC) grouping for the quadrupole time series in m12b, based on mSSA decomposition. \textit{Top row}: PC amplitudes as a function of time (half-window length shown). The left panel presents PC0–PC3, which exhibit similar long-timescale evolution, while the right panel shows PC4–PC11, dominated by higher-frequency variations. \textit{Bottom row}: the left panel displays the DFT power spectrum for each PC, highlighting characteristic timescales; the right panel shows the w-correlation matrix, quantifying the correlation between different PCs. These analyses define two primary groupings—one associated with filamentary structure (PC0–PC3) and another with the LMC-analog response (PC4–PC11). PC12, which exhibits low correlation and significantly lower power, remains ungrouped.}
\label{fig:mssa_details_quad}
\end{figure*}

In this appendix, we detail the methodology used to group PC from the SVD of the lagged trajectory matrix in mSSA. Since the input coefficient time series are often interdependent, a single dynamical signature can be spread across multiple PCs. Additionally, varying frequency modes may be distributed among different PCs, necessitating a systematic approach to identifying physically meaningful groupings.

The number of PCs required to describe a dynamical signal is tied to its temporal complexity. For instance, linear or exponential trends (e.g., gradual filamentary growth) are captured by a single PC, as their evolution depends only on the immediate past. Periodic signals (e.g., satellite orbital motion), however, require groups of PCs to encode amplitude and phase, analogous to sine and cosine components. Nonstationary signals---such as those involving abrupt changes (e.g., pericenter passages) or frequency modulation---demand additional PCs to account for their evolving structure.

One of the most fundamental yet tedious methods for grouping PCs involves reconstructing gravitational fields from individual PCs and visually inspecting their spatial and temporal behavior. However, a more quantitative grouping can be achieved by applying the following criteria:

\begin{itemize}
    \item \textbf{\textit{Singular Values.}} PCs associated with the same dynamical process tend to have comparable singular values.
    
    \item \textbf{\textit{Time Evolution.}} PCs that exhibit similar time-dependent behavior can be identified by plotting their amplitudes over time (top panels of Fig.~\ref{fig:mssa_details_quad}).

    \item \textbf{\textit{Frequency Analysis.}} Applying a discrete Fourier transform (DFT) to the PCs allows us to examine their dominant frequency components and identify those that share similar power spectra (bottom left panel of Fig.~\ref{fig:mssa_details_quad}).

    \item \textbf{\textit{w-Correlation Matrix.}} The weighted correlation matrix, as introduced in \citet{weinberg2021using}, quantifies the correlation between different PCs, helping to establish natural groupings  (bottom right panel of Fig.~\ref{fig:mssa_details_quad}).   
\end{itemize}

For the case of m12b, we analyze the quadrupole ($\ell=2$) coefficients, including all azimuthal modes ($m=0,1,2$) and the full range of radial orders ($n$). We focus on the first 12 PCs, as they collectively account for 98\% of the total variance, with the singular values dropping exponentially beyond this set. In our analysis, the filament-driven quadrupole aligns with a quasi-linear trend, requiring few PCs, while the satellite-induced signal spans multiple PCs due to its quasi-periodic orbital motion and transient interactions with the halo.

Fig.~\ref{fig:mssa_details_quad} illustrates the process used to define PC groupings. The top row presents the PC amplitudes (color-coded) over time. The left panel shows the four most dominant PCs (PC0–PC3), while the right panel displays PC4–PC11. These groupings are made based on their temporal signatures and amplitudes. The bottom row provides further insight: the left panel shows the DFT power spectrum for each PC, revealing characteristic timescales, while the right panel displays the w-correlation matrix, quantifying the correlation between different PCs. From this analysis, we identify two primary groupings:

\begin{itemize}
    \item \textbf{Filamentary structure group}: PC0–PC3 exhibit low-frequency variations with characteristic timescales of $\sim4$ Gyr, consistent with the slow evolution of the large-scale halo anisotropy. These components also share power at a secondary $\sim3$ Gyr timescale. The w-correlation matrix confirms their mutual correlation.

    \item \textbf{LMC-analog response group}: PC4–PC11 contain power at higher frequencies, peaking at $\sim1.5$ Gyr, indicative of the satellite-induced perturbations. These PCs show moderate correlations in the w-correlation matrix.

    \item \textbf{Uncorrelated component:} As an example, PC12 exhibits three-orders-of-magnitude-lower power at very high frequencies and appears uncorrelated with other PCs, suggesting that it primarily represents noise or a higher-order transient effect.
\end{itemize}

In Sec.~\ref{sec:quad_mssa}, we utilize these identified filamentary and LMC-associated frequency groups to reconstruct and analyze the gravitational structures driving halo evolution.

\setlength{\bibsep}{0pt} 
\renewcommand{\baselinestretch}{1.0} 
\bibliography{ref}{}
\bibliographystyle{aasjournalv7}

\end{document}